\documentclass{article}

\usepackage{PRIMEarxiv}

\usepackage[utf8]{inputenc} 
\usepackage[T1]{fontenc}    
\usepackage{hyperref}       
\usepackage{url}            
\usepackage{booktabs}       
\usepackage{amsfonts}       
\usepackage{nicefrac}       
\usepackage{microtype}      
\usepackage{lipsum}
\usepackage{fancyhdr}       
\usepackage{graphicx}       
\graphicspath{{media/}}     
\usepackage{graphicx}
\usepackage{subfigure}
\title{PM-Gati Shakti: Advancing India's Energy Future through Demand Forecasting - A Case Study

}

\author{
  SujayKumar Reddy M\\
  CSE Department \\
  Vellore Institute of Technology ,Vellore \\
  \texttt{sujaykumarreddy.m2020@vitstudent.ac.in} \\
   \And
  Gopakumar G \\
  CSE Department \\
  National Institute of Technology, Calicut \\
  \texttt{gopakumarg@nitc.ac.in} \\
}

\begin{document}
\maketitle

\begin{abstract}

PM-Gati-Shakti Initiative, integration of ministries, including railways, ports, waterways, logistic infrastructure, mass transport, airports, and roads. Aimed at enhancing connectivity and bolstering the competitiveness of Indian businesses, the initiative focuses on six pivotal pillars known as "Connectivity for Productivity": comprehensiveness, prioritization, optimization, synchronization, analytical, and dynamic. In this study, we explore the application of these pillars to address the problem of "Maximum Demand Forecasting in Delhi." Through a detailed case study, we seek to comprehend and formalize the use cases associated with this crucial forecasting task, illuminating the potential and impact of the PM-Gati Shakti scheme in shaping India's energy landscape and driving sustainable growth. Electricity forecasting plays a very significant role in the power grid as it is required to maintain a balance between supply and load demand at all times, to provide a quality electricity supply, for Financial planning, generation reserve, and many more. Forecasting helps not only in Production Planning but also in Scheduling like Import / Export which is very often in India and mostly required by the rural areas and North Eastern Regions of India. As Electrical Forecasting includes many factors which cannot be detected by the models out there, We use Classical Forecasting Techniques to extract the seasonal patterns from the daily data of Maximum Demand for the Union Territory Delhi. This research contributes to the power supply industry by helping to reduce the occurrence of disasters such as blackouts, power cuts, and increased tariffs imposed by regulatory commissions. The forecasting techniques can also help in reducing OD and UD of Power for different regions. We use the Data provided by a department from the Ministry of Power and use different forecast models including Seasonal forecasts for daily data.
\end{abstract}


\section{Introduction}

The Indian Energy GRID is maintained by POWERGRID \cite{POWERGRID} which has the objectives of running the GRID efficiently and installing transmission lines etc... and the second one is the National Load Dispatch Center (NLDC) \cite{grid} which concentrates on Supervision over the Regional Load Dispatch Centres, Scheduling and dispatch of electricity over inter-regional links in accordance with grid standards specified by the Authority and grid code specified by Central Commission in coordination with Regional Load Dispatch Centres, Monitoring of operations and grid security of the National Grid, etc... This research mainly focusses on this NLDC which is a Division of Ministry of Power. 
\hfill \break

\begin{figure}[t]
\includegraphics[scale=0.6]{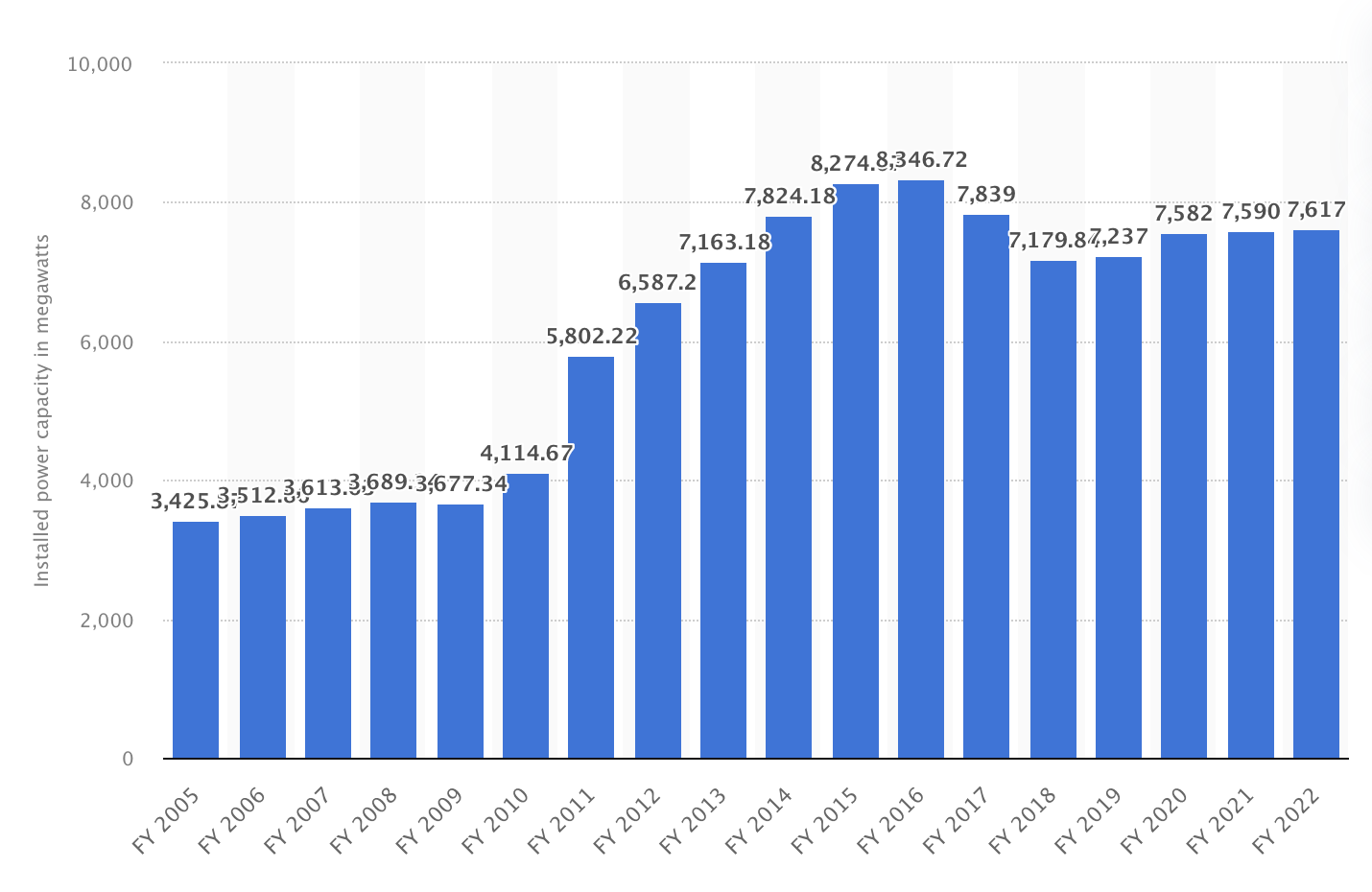}
\includegraphics[scale=0.7]{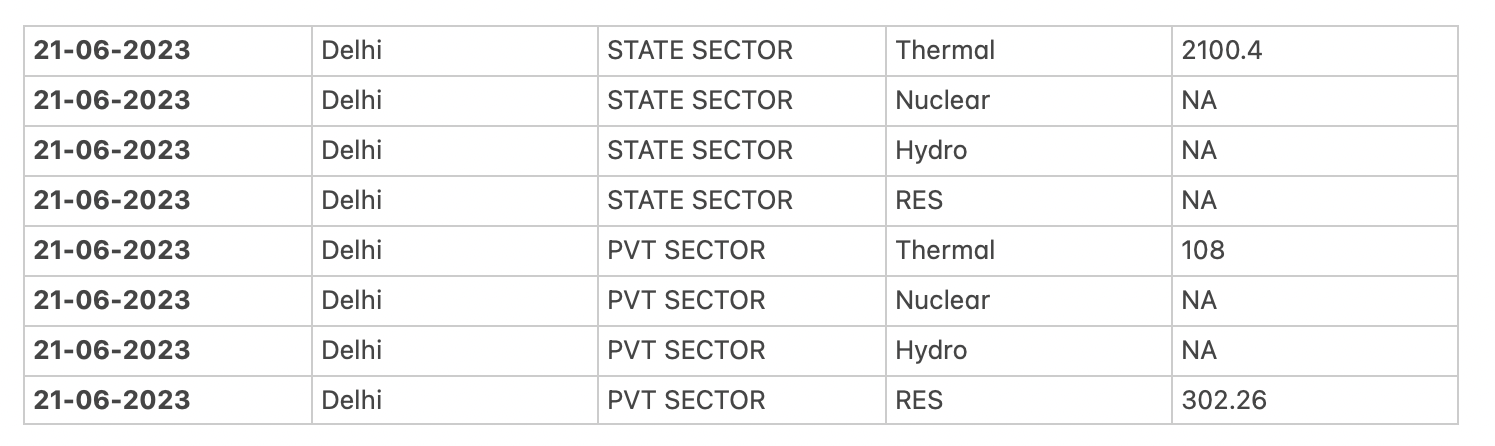}
\caption{Installed Capacity}
\centering
\end{figure}
\begin{figure}[htbp]
  \centering

  \subfigure[Per-capita Availability of Power]{\includegraphics[width=0.4\textwidth]{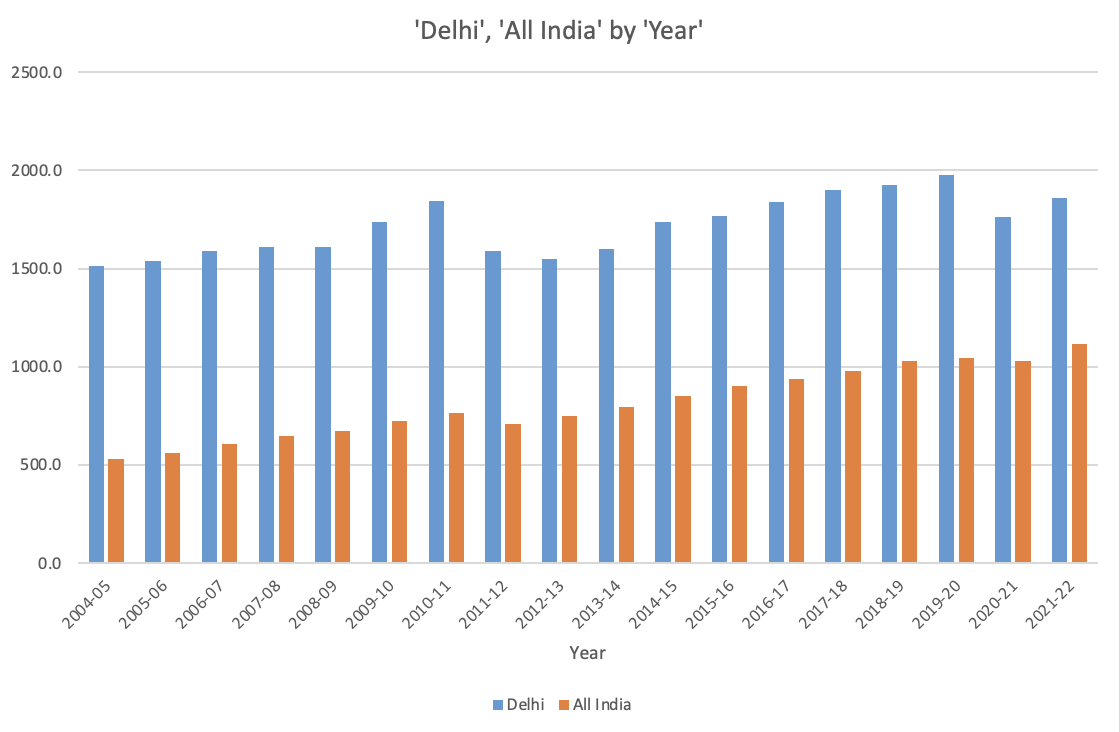}}
  \hfill
  \subfigure[Availability of Power]{\includegraphics[width=0.4\textwidth]{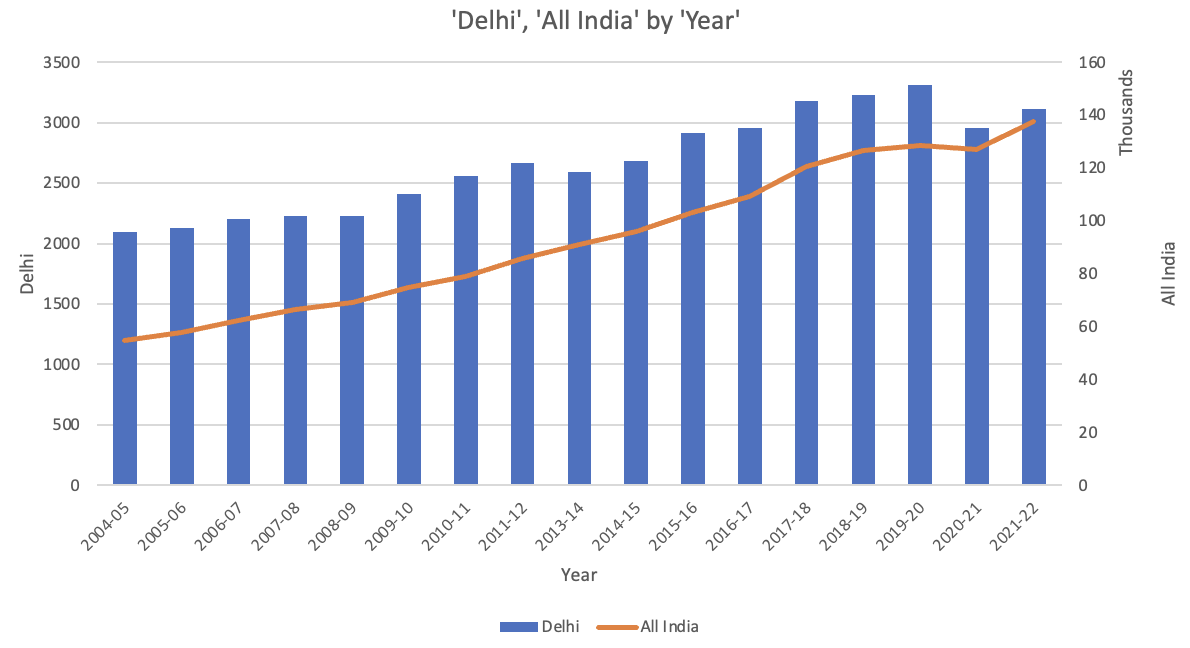}}
  
  \vspace{0.5cm}
  
  \subfigure[Total Installed Capacity of grid Interactive Renewable Power]{\includegraphics[width=0.4\textwidth]{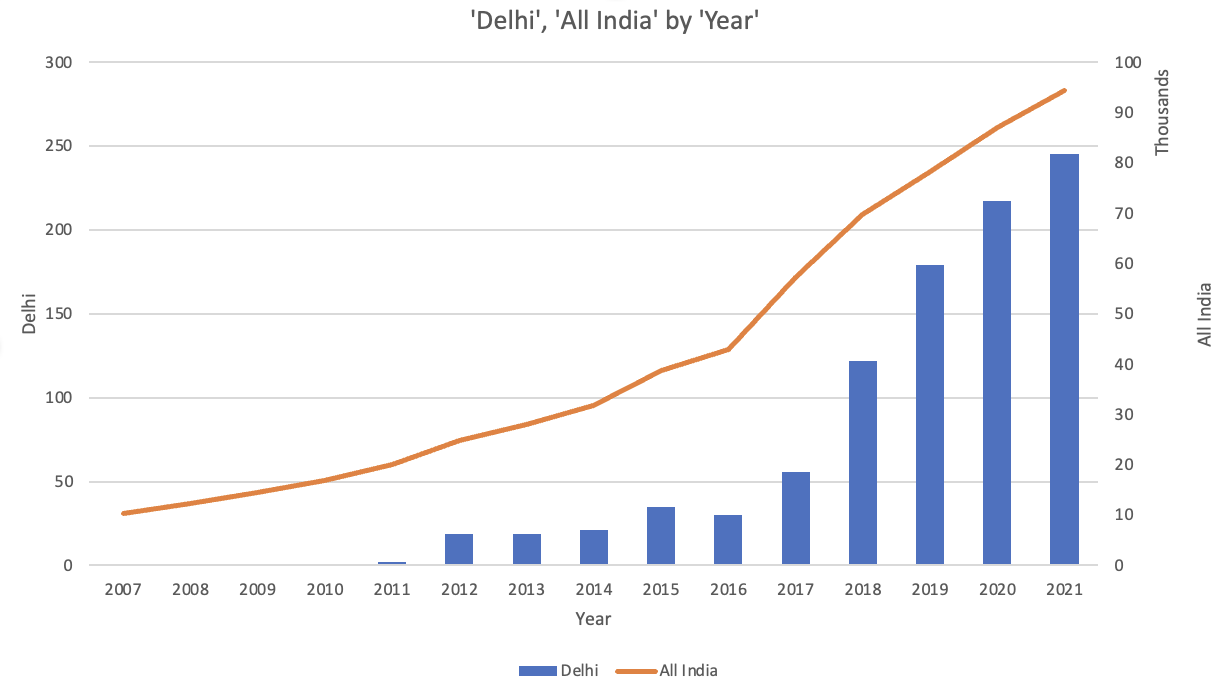}}
  \hfill
  \subfigure[Power Requirement]{\includegraphics[width=0.4\textwidth]{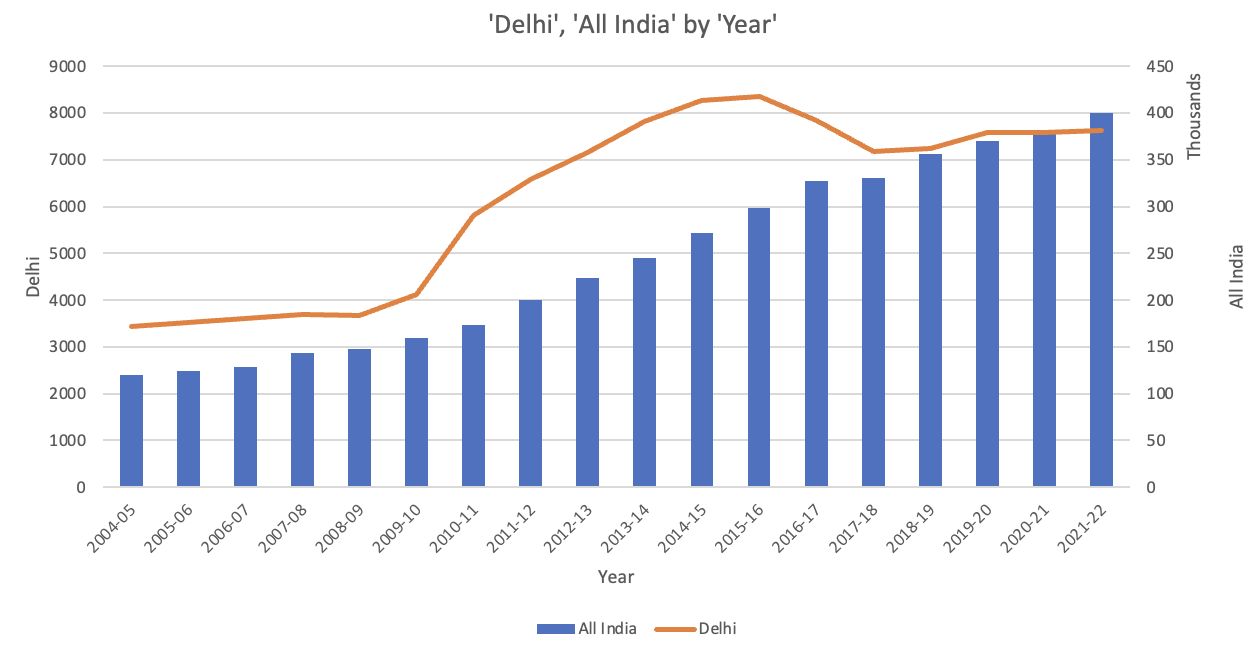}}

  \caption{Statistics on the Delhi's Power}
  \label{fig:collage}
\end{figure}

Fig 1 \cite{rbi} describes the Yearly Installed Power Capacity in Delhi. The highest installed Capacity was 8,346.72 MW in Fiscal year 2016. which is responsible for sending the energy from Stations to sub-stations and to discuss and then to homes, industries, commercials, etc... As of 21-06-2023 The installed Capacity Sector-wise data \cite{data} gives an overview of what type of Thermal Plants which are present in Delhi and also what types of Energy resources are present in Delhi. There are 11 Thermal Stations in Delhi with 4 400KV Substations and 42 200 KV Substations \cite{DTLSubstations}. 
\hfill \break

\subsection{Delhi Yearly Power Statistics}

To further Analyse Delhi's Power, this paper uses the \cite{rbi} RBI's handbook contains yearly state-wise data to analyze the Delhi v/s Whole India's data. Fig 2(a) gives Delhi v/s whole India's Statewise Per-capita of Power raises to 1974.4 Kilo-Watt Hour in Delhi in 2018-2019 and in India 1115.3 Kilo-Watt Hours in 2021-2022. Fig 2(b) gives an Availability of Power at Delhi raises to 3308 net Crore Units in 2019-2020 and 137402 net crore units in 2021-2022 for All India, Fig 2(c) gives the Total Installed Capacity of RES Power raises to 245 Mega Watt at 2021 for Delhi and 94434 MegaWatt at 2021 for all India. Fig 2(d) depicts the Power Requirement for Delhi has the maximum reading at 3309 net crore units in 2019-2020 and for all of India 137981 net crore units in 2021-2022. 

\subsection{Relation to Demand Forecasting}
\includegraphics[scale=0.8]{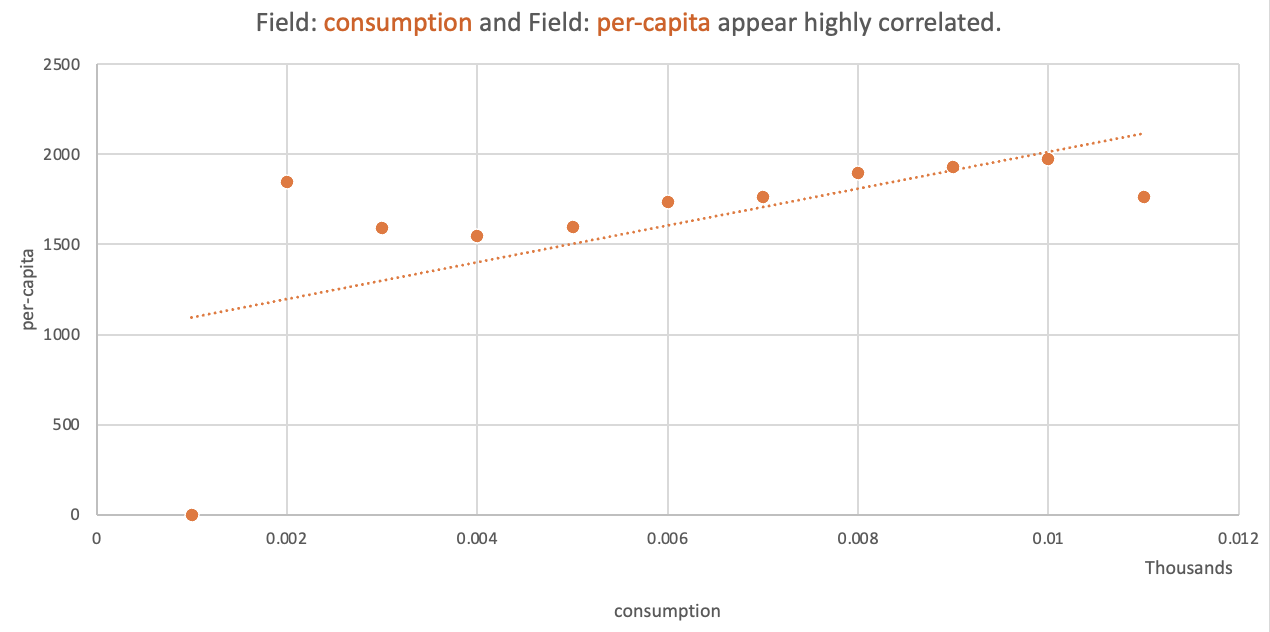}
The correlation between these variables with the Consumption of Electricity of Delhi \cite{delhi} Now, as we speak Consumption by the consumers is based on the Maximum Demand which has been recorded per day in the National Load Dispatch Center (NLDC) As all these entities are highly correlated we push our limits to only use univariant analysis of Maximum Demand attained in the Daily data produced by the Delhi Consumers.

\subsection{PM-Gati Scheme Initiative}
Statistically speaking, many papers have demonstrated that Economic Factors are valuable data for Maximum Demand Forecasting. Although these findings have been successful, we present a baseline model focusing solely on one variable to capture the Auto-Correlations between days, weeks, months, and even years. This approach aims to create baseline performance models due to Seasonal Dependence. To achieve this, we employ various pre-processing techniques and utilize a selection of Machine Learning and Time Series Forecasting models that have been previously applied to create our Baseline model.

The "Integration of Ministries" initiative combines seven drivers: railways, ports, waterways, logistic infrastructure, mass transport, airports, and roads. The primary goal is to enhance connectivity and boost the competitiveness of Indian businesses. This integration is known as "Gati-Shakti," and it rests on six pillars of Connectivity for Productivity.
\begin{enumerate}
    \item Comprehensiveness
    \item Prioritization
    \item Optimization
    \item Synchronization
    \item Analytical
    \item Dynamic
\end{enumerate}

Given these Pillars, we try to formalize and study the use cases which are provided for this problem Statement, "Maximum Demand Forecasting in Delhi", Here's a case study to understand ... 
\hfill \break

Let us say that the Ministry of Steel\cite{press} created an initiative to increase domestic production that would lower dependence on imported Steel and would result in considerable savings of foreign exchange. As this seems reasonable, there are some pros and cons related to this one being less Foreign Exchange obviously and a con being Increased Electricity Consumption. Data explains from the Ministry of Coal to the Ministry of Power to increase the installed Capacity or Power Generation can be captured through Comprehensiveness.

As there's an increase in demand for Electricity, the Ministry of Power tries to solve this problem but as the transparency of Ministries in PM-Gati-Shakti provides to consolidate the increase in demand like the Ministry of Railways tries to optimize routes on weekends and majorly metro trains which are in Delhi can be optimized to consolidate this increase in demand which can be captured through Prioritization, Optimization while maintaining a Holistic Approach. Note that this is a scenario that can be predicted but there will be many scenarios as possible in-order to facilitate the Maximum Demand Forecasting.

\section{Related Work}


This Section consists of Background analysis a.k.a Literature review of the Demand Forecasting Techniques adopted and Analyzing the recent reports by the Ministries of Government. This section focuses on many different attributes which are needed to be considered by the previous research done by the individuals.

\subsection{Deep Learning}
\begin{figure}[htbp]
  \centering

  \subfigure[Deep Learning Papers Analysis]{\includegraphics[width=0.4\textwidth]{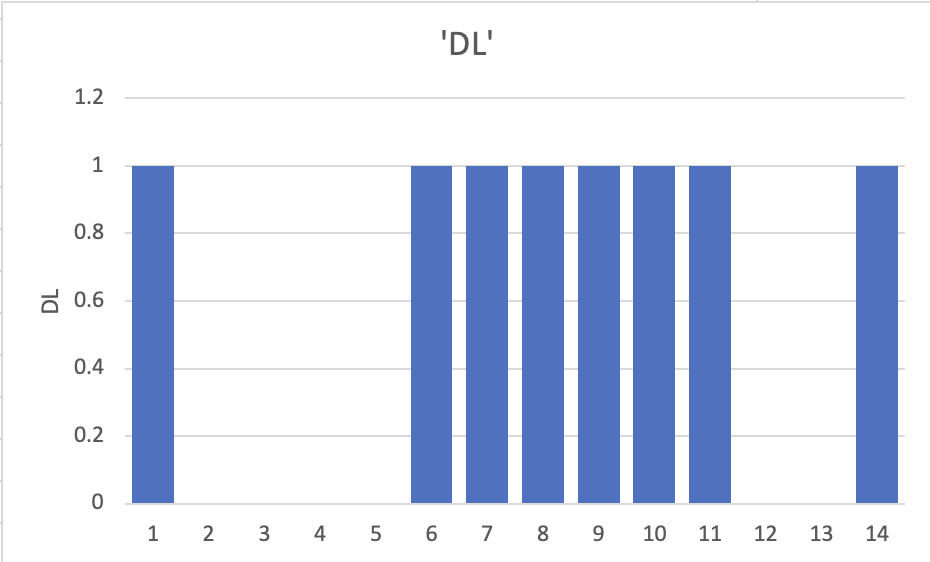}}
  \hfill
  \subfigure[Machine Learning Papers Analysis]{\includegraphics[width=0.4\textwidth]{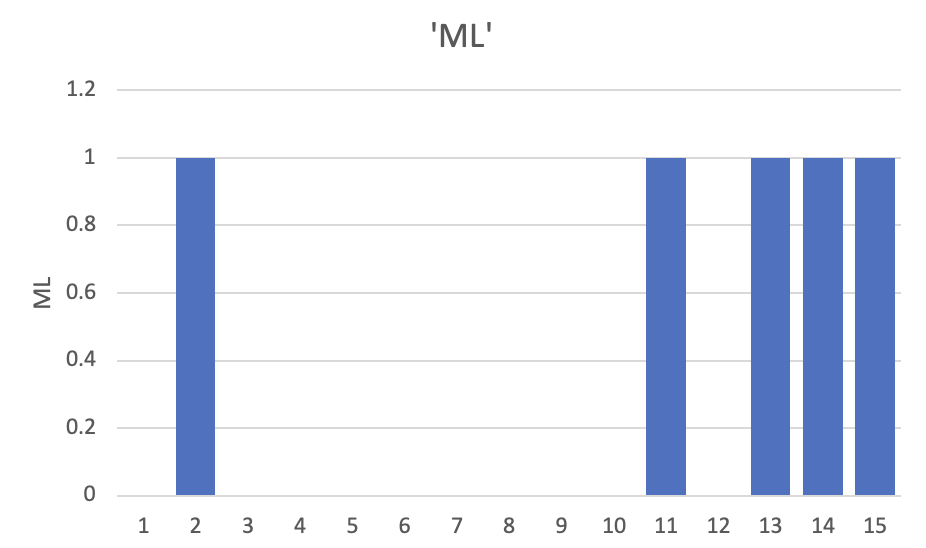}}
  
  \vspace{0.5cm}
  
  \subfigure[Regression based Papers Analysis]{\includegraphics[width=0.4\textwidth]{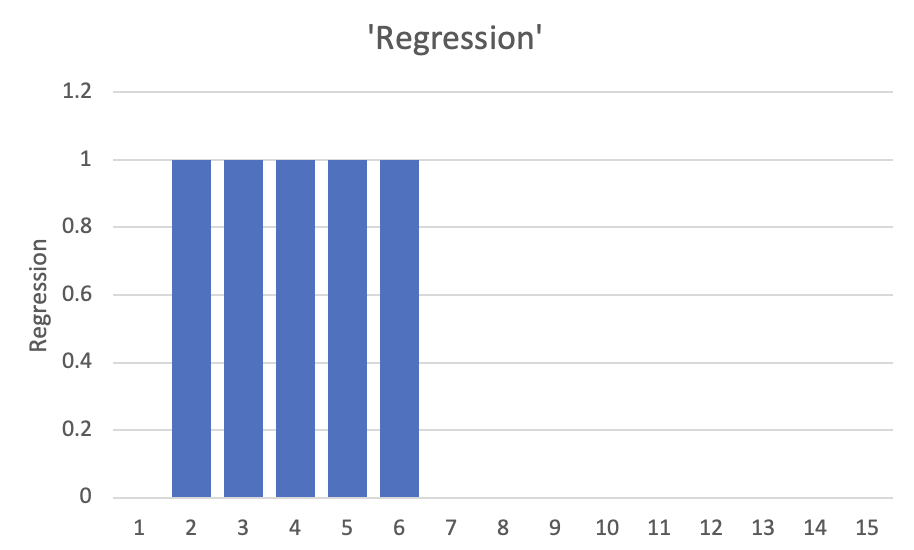}}
  \hfill
  \subfigure[Predictors used in the Dataset Papers]{\includegraphics[width=0.4\textwidth]{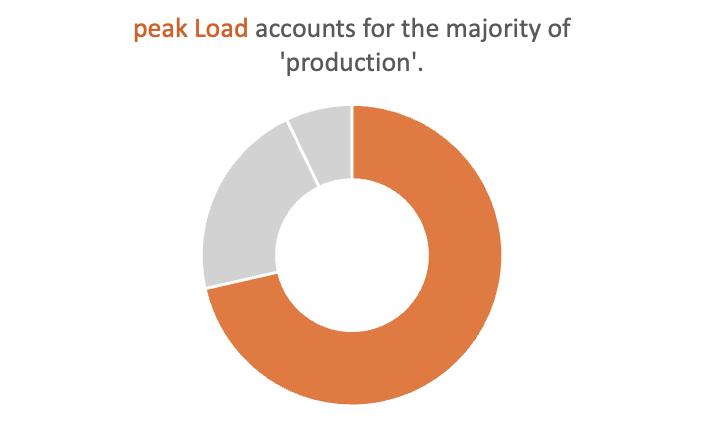}}

  \caption{Literature Survey Framework Analysis}
  \label{fig:collage}
\end{figure}

\hfill \break

Fig 3 (a) gives us a clear understanding of a number of Deep Learning papers which are mostly based on Indian Authors published in IEEE. 
Anil et al \cite{d1} use the Levenberg-Marquardt back propagation algorithm ANN on the day ahead Short term load Forecasting on the state of Uttar Pradesh trained on hourly data with the MAPE score of average MAPE 3.05, This work suggests using the ANN model to check with our dataset too. Navneet et al \cite{d2} use the New Delhi ADEL data to forecast the load by using different Neural Network Architectures in which ELMANN Neural Network Architecture has given good accuracy.
Dharmoju et al \cite{d3} provided a sector of Residental buildings by the United States Dataset by using LSTM (Long Short Term Memory) model for monthly forecasting. Shaswat et al \cite{d4} use a Temporal Fusion Architecture to capture the interactions which are scaled between 0 and 1 for daily data which achieves 4.15\% more than the existing models and this is for the whole of India which is not region specific. Saravanan et al \cite{d5} use economic factors like GDP, national income, consumer price index, etc.. with that they used Principle component Analysis followed by ANN which gives the highest accuracy of MAPE score of 0.43. Vishnu et al \cite{d6} concentrate on the work on Renewable Energy Resources and devised two major LSTM (Long Short Term Memory) models. 

\subsection{Machine Learning}
 Fig 3 (b) gives us a clear understanding of a number of Machine Learning papers and their analysis in all the papers (as ids). Christos et al \cite{m1} which also forecasts the Peak Demand in the electrical sector of the Producers side. They used a Netherland dataset from 3 regions and analyzes comparatively with all the existing Forecasting methods such as ARIMA, Ridge, and Lasso Regression and the results suggested that the Bi-directional LSTM. Saravanan et al \cite{m2} formalize a set of 64 if-else statements and the variables include per-capita GDP, population, and Import/Export and they have achieved the MAPE of 2.3. Mannish et al \cite{m3} devised an Ensemble Approach for the DISCOMs in the region Delhi for the Post Covid Scenario and their proposed model includes combining XGBoost, LightGBM, and CatBoost algorithms which achieved an average MAPE of 5.0. Banga et al \cite{m4} compared many Machine Learning Algorithms for the dataset which considers 29 attributes and the Facebook Prophet model outscores daily and hourly datasets with MAPE scores of 0.4 and 0.2.

\subsection{Regression Based Learning}
Fig 3 (c) gives us a clear understanding of a number of Regression papers or Auto Regression papers compared with all the papers (as ids). note these papers are based completely on uni-variant/single-variant datasets. Carlos et al \cite{r1} analyzed a time series dataset for Brazilian Electricity Demand Forecasting and divides Brazil into 2 regions and forecasts the Electricity demand according by using ARIMA models. Kakoli et al \cite{r2} forecasted the electricity demand for the state Assam in the Northeast Region and the results suggest that to use the Seasonal ARIMA model with the formula given below SARIMA(0,1,1)(0,1,1,7) with the MAPE of 10.7. Srinivasa et al \cite{r3} provided a forecasting method that is formulated monthly for the whole of India without considering the states and regions. It has been found that the MSARIMA model outperforms CEA forecasts in both in-sample static and out-of-sample dynamic forecast horizons in all five regional grids in India.  

\hfill \break

\subsection{Summary of Literature Survey}

\begin{center}
\begin{tabular}{|c|c|}
  \hline
  \textbf{Clusters } & \textbf{Paper Citations}  \\
  \hline
  Machine Learning & \cite{m1} \cite{m2} \cite{m3}  \cite{m4} \\
  \hline
  Deep Learning  &   \cite{d1} \cite{d2}  \cite{d3} \cite{d4}  \cite{d5} \cite{d6} \\
  \hline
  Auto-Regressive Models & \cite{r1} \cite{r2} \cite{r3} \\
  \hline
\end{tabular}
    
\end{center}

\section{Methodology}

\subsection{Data Overview}
The Features of the data which has been provided below
\begin{enumerate}
    \item Date (DD/MM/YYYY)	
    \item Max.Demand met during the day (MW)	
    \item Shortage during maximum Demand (MW)	
    \item Energy Met (MU)	
    \item Drawal Schedule (MU)	
    \item OD(+)/UD(-) (MU)	
    \item Max OD (MW)	
    \item Energy Shortage (MU)
\end{enumerate}

Energy Shortage (MU) feature is not available every day. This feature is recorded from 2017-05-09 as per the reports generated by the PSP by POSOCO. The data is available here \cite{dataset}. This paper focuses on Univariant Analysis not to make it as complex, but to consider Max Demand met during the day (MW) as a single column. 
\hfill \break

As the column "Max Demand met During the day" is the major feature that we consider. The data starts from  2013-04-01 to 2023-05-31 which consists of 3713 days but the data points only consist of 3640 with the missing data we use the Imputation Techniques dataset and Non-Imputation Techniques dataset (no\_null).
\hfill \break

For Imputation Techniques, this paper considers Mean, Median, Mode, and Linear Interpolation Imputation data that has been imputed. So, as combined this generates 5 datasets where the models are applied to compare the performance of which Imputation is good.
\hfill \break

\begin{table}[]
    \centering
    
    \begin{tabular}{|c|c|}
    \hline
        1 &  dropna-dataset\\
         \hline
         2 & mean Imputation dataset\\
        \hline
        3 &  median Imputation dataset\\
         \hline
         4 & mode Imputation dataset\\
         \hline
         5 & linear-Interpolation Imputation dataset\\
         \hline
    \end{tabular}
    \caption{ Datasets names which are created from the reports}
    \label{tab:my_label}
\end{table}

Table 1 depicts the datasets which are created from the univariant data taken from one of the features in the dataset. ("Max. Demand met during the day (MW) "). As explained in the Methodology section the dataset is divided into train and test to take the MAPE score. For ARIMA models we generate train MAPE and test MAPE to check the overfitting criteria also with AIC, BIC, log(p), etc...

\section{Forecasting Models}

This paper develops a Time series forecasting model ARIMA which is known to be an Auto-Regressive Integrated Moving Average. the models which include AR, MA and ARMA, and ARIMA, and develop a model list from these regression types using the parameters. The major parameters included in the arima model are p, q, and d where p is the parameter for Auto-regressive co-efficient which says about how many days have the co-relation between today's date.

The real-world data tends to be always non-Stationary. A signal is said to be stationary if its statistical properties like mean, standard deviation, trend, etc... doesn't change over time. To check if the time series is stationary or not, we use Augmented-Dickey Fuller Test where the null hypothesis is "the time series contains a unit root and is non-stationary". The results of the Augmented Dickey-Fuller Test for each of the Imputation datasets are given in the below sections.

\subsection{Datasets Analysis and ADF test results}
The Figure below gives a plot of the whole dataset without dividing into train and test.
\includegraphics[scale=0.5]{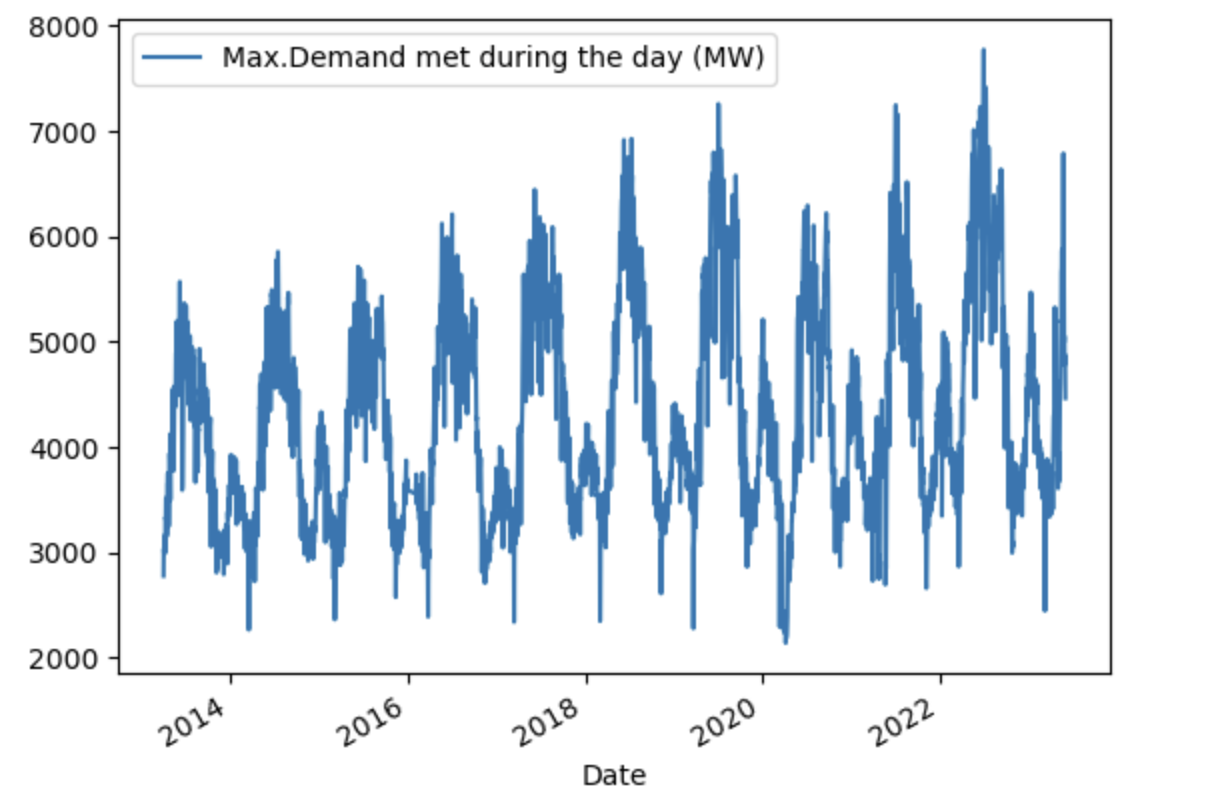}

The ADF test for no-imputation dataset test-statistic = -5.45 p-value = 2.55e-06 for no-imputation dataset for first difference to try to change the time-series to stationary. test-statistic = -10.403 p-value = 1.88e-18 for Second Difference, which is not suggested as the p-value is zero (over-differencing) test-statistic = -21.152  p-value = 0.0

\hfill \break

This figure below depicts the Mean Imputation dataset Plot. \\

\includegraphics[scale=0.5]{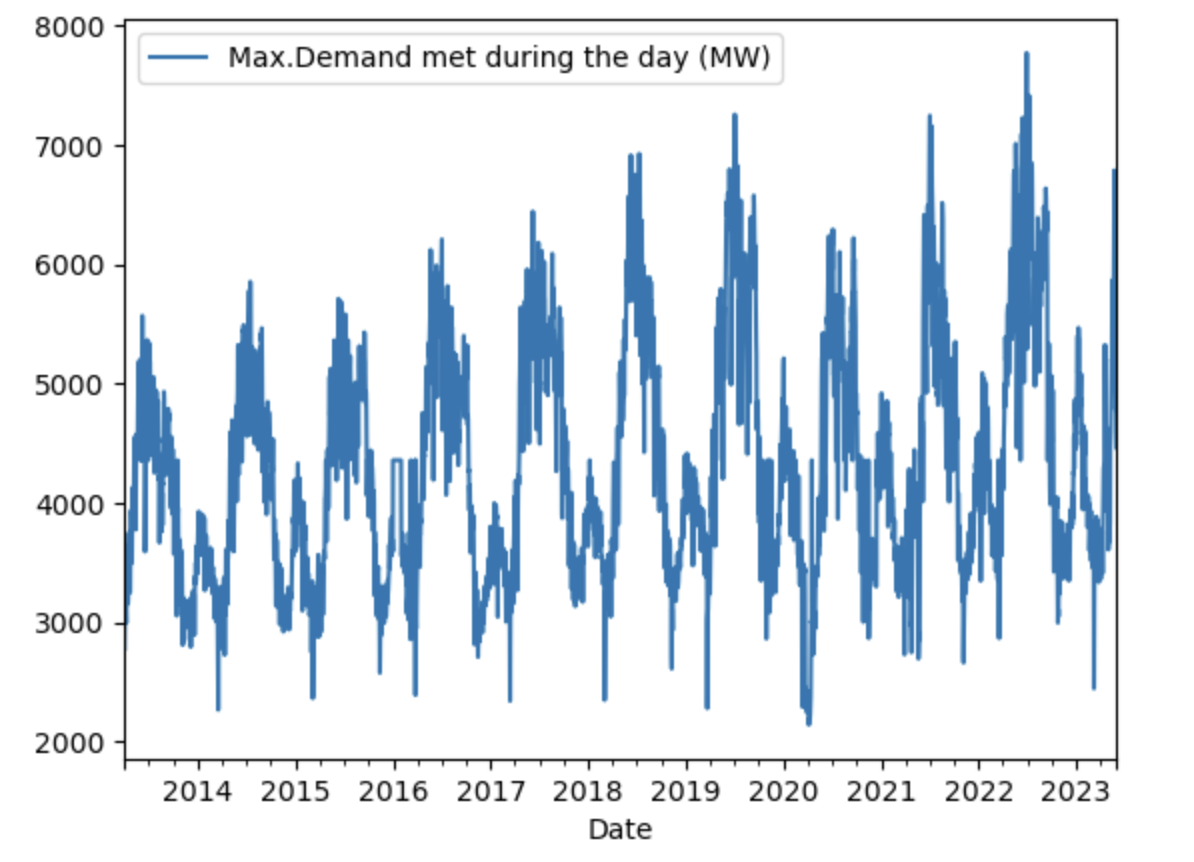}

The ADF test for Mean imputation dataset test-statistic = -5.393  p-value = 3.49e-06 for first difference to try to change the time-series to stationary. test-statistic = -10.073 p-value = 1.23e-17 for Second Difference, which is not suggested as the p-value is zero (over-differencing) test-statistic = -21.617 
p-value = 0.0

\hfill \break

This figure below depicts the Median Imputation dataset Plot. \\
\includegraphics[scale=0.5]{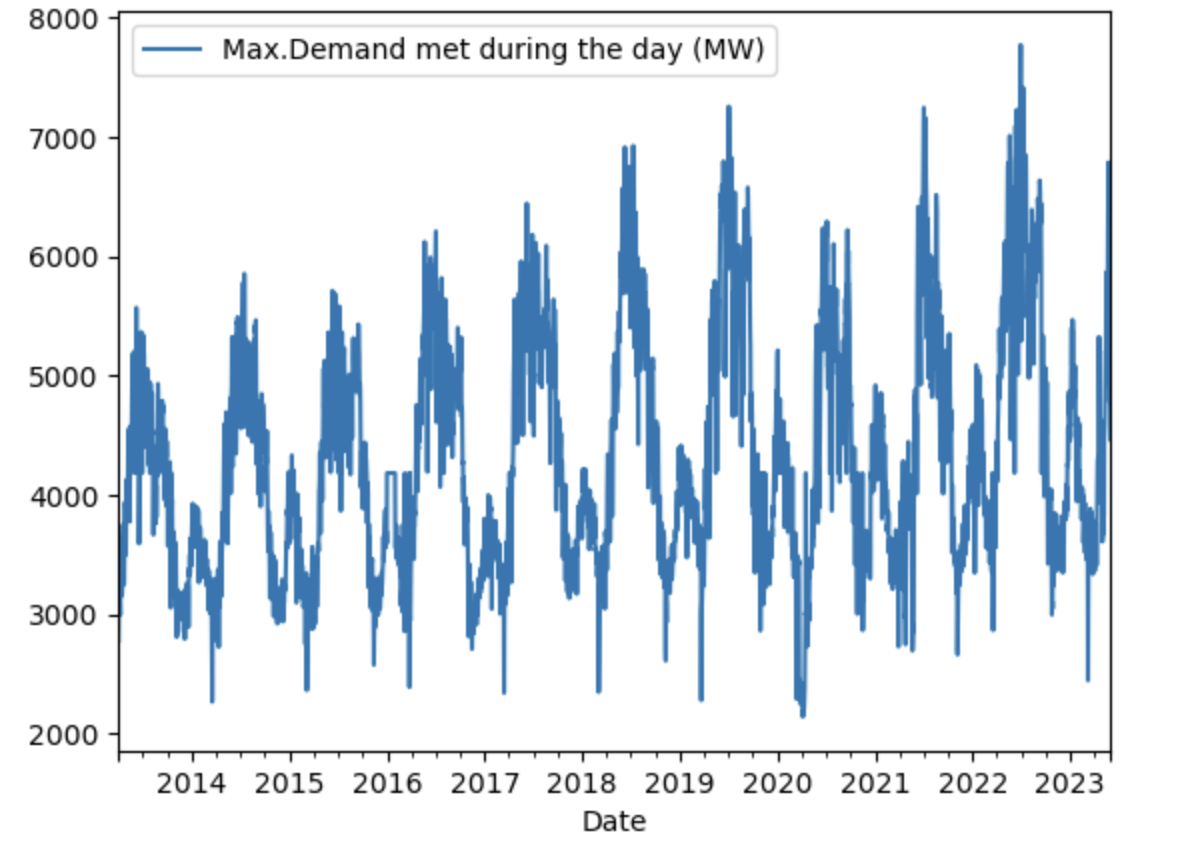}

The ADF test for Median imputation dataset test-statistic = -5.363 p-value = 4.042e-06, for first difference to try to change the time-series to stationary. test-statistic = -10.075 p-value = 1.223e-17 for Second Difference, which is not suggested as the p-value is zero (over-differencing) test-statistic = -21.686 p-value = 0.0

This figure below depicts the Mode Imputation dataset Plot. \\
\includegraphics[scale=0.5]{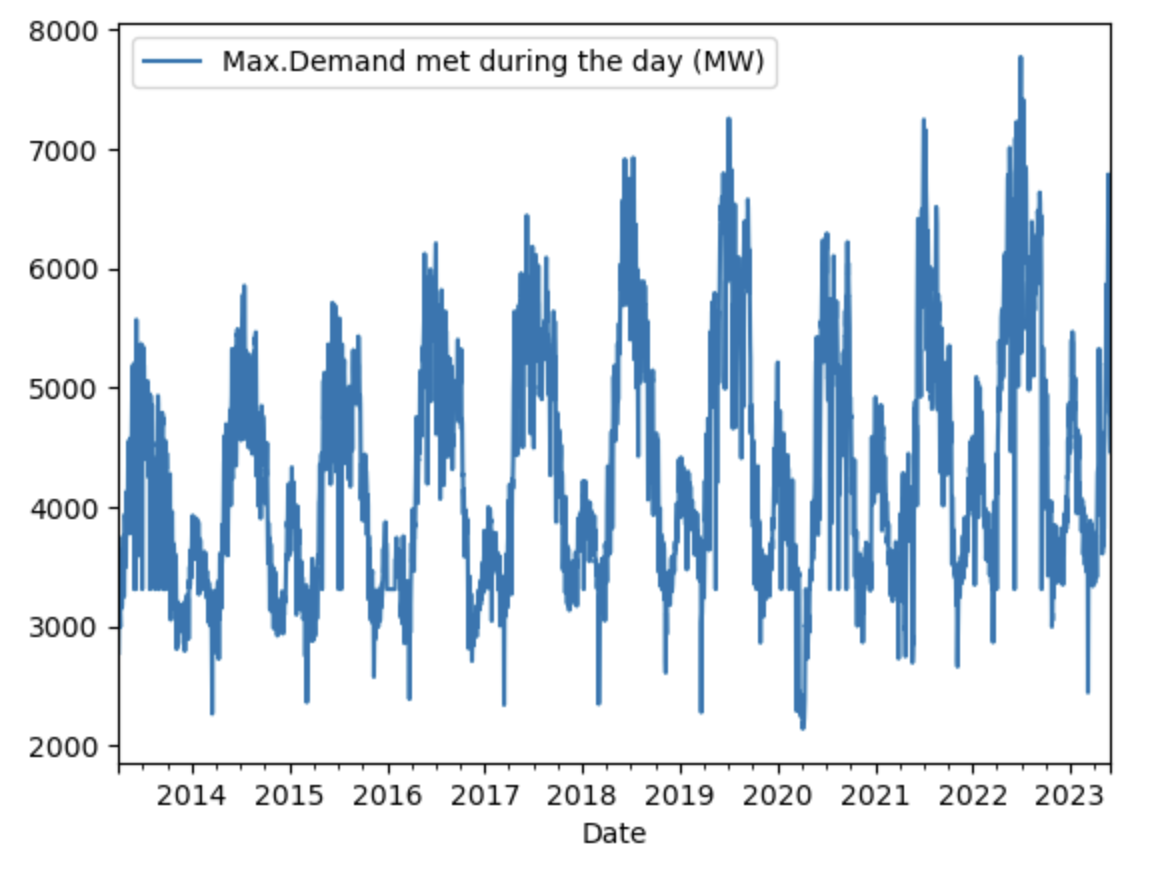}

The ADF test for Mode imputation dataset test-statistic = -5.227 p-value = 7.73e-06 for first difference to try to change the time-series to stationary. test-statistic = -10.258 p-value = 4.31e-18 for Second Difference, which is not suggested as the p-value is zero (over-differencing) test-statistic = -22.121 p-value = 0.0

This figure below depicts the Linear Interpolation Imputation dataset Plot. \\
\includegraphics[scale=0.5]{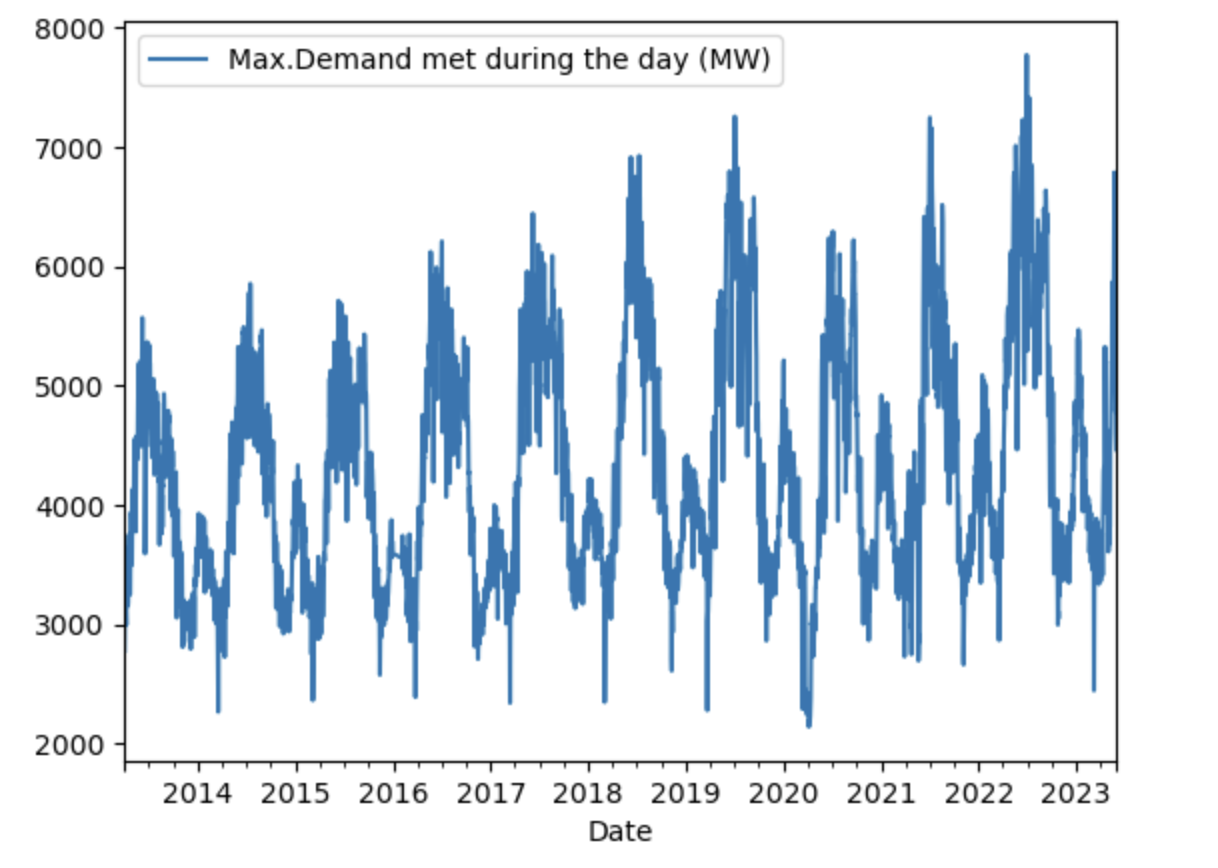}

The ADF test for Linear Interpolation imputation dataset 
test-statistic = -5.390 p-value = 3.53e-06 for first difference to try to change the time-series to stationary test-statistic = -10.072   p-value = 1.24e-17 for Second Difference, which is not suggested as the p-value is zero (over-differencing) test-statistic = -21.44 p-value = 0.0

\section{Analysis using ACF and PACF plots}

\subsection{No Imputation}

The Auto-Correlation Function and Partial Auto-Correlation Function Graph for the original dataset
\hfill \break
\includegraphics[scale=0.4]{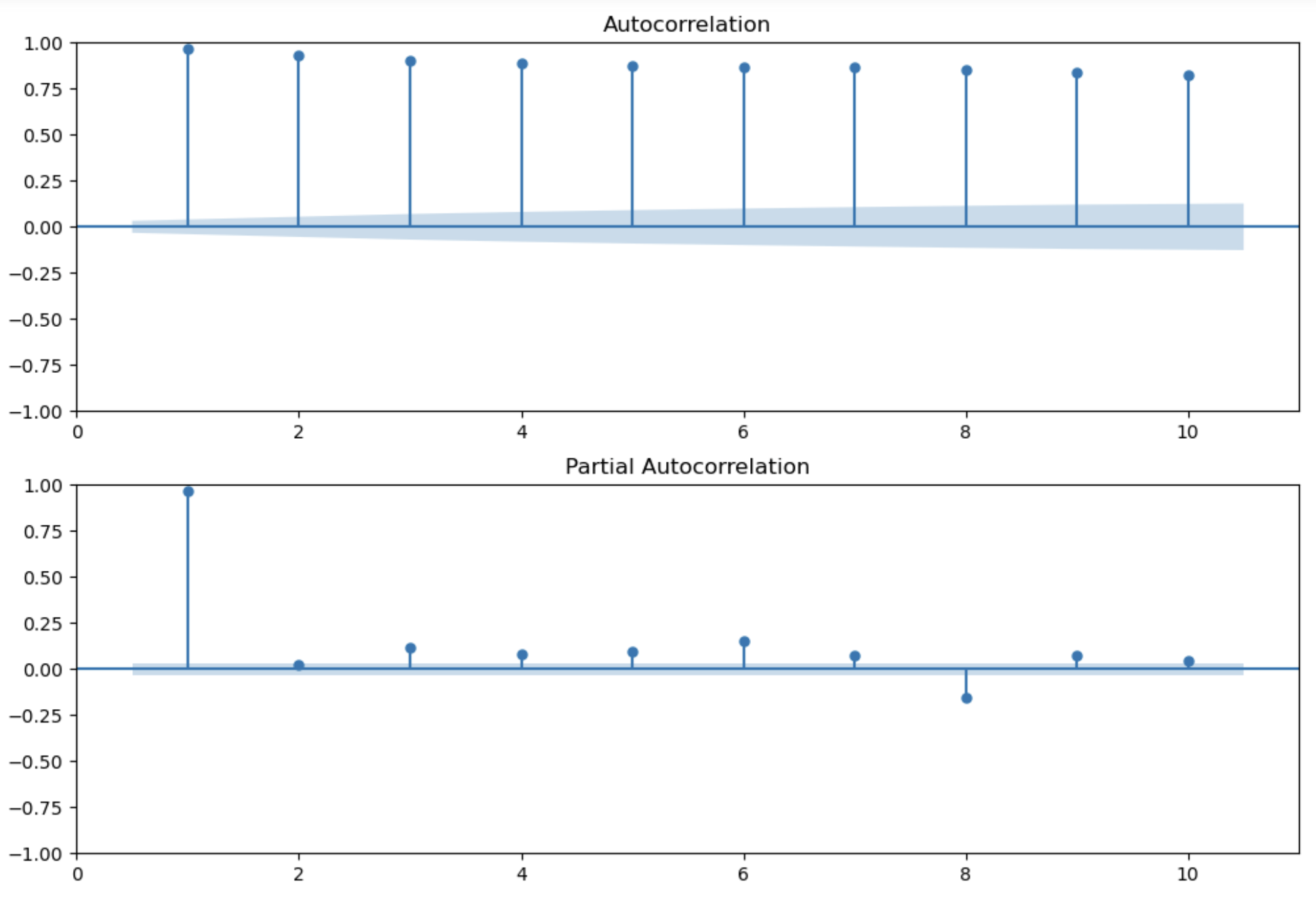}

The ADF test for first difference to try to change the time-series to stationary.
\hfill \break
\includegraphics[scale=0.4]{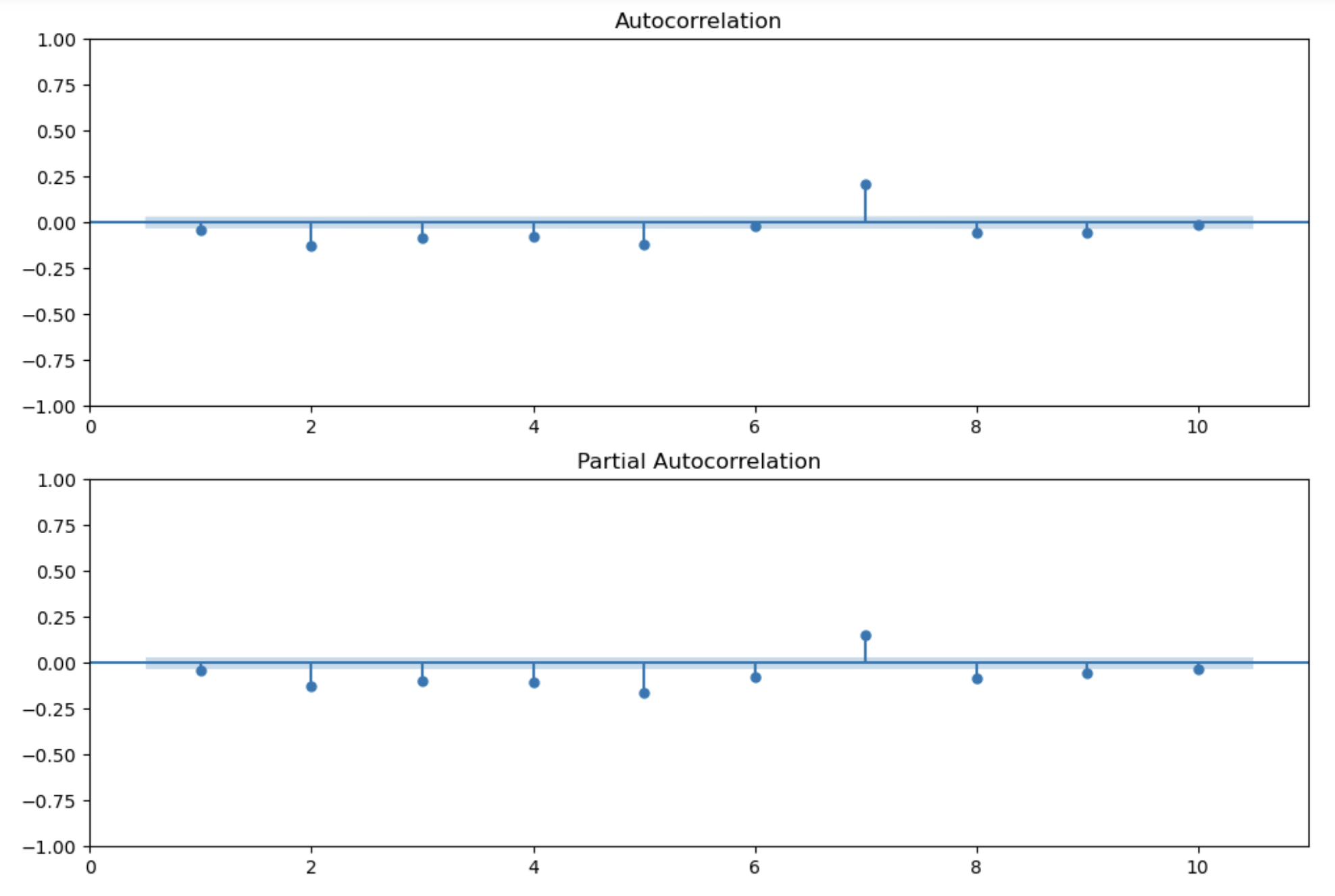}

\subsection{Mean Imputation}
The Auto-Correlation Function and Partial Auto-Correlation Function Graph for the original dataset
\hfill \break
\includegraphics[scale=0.4]{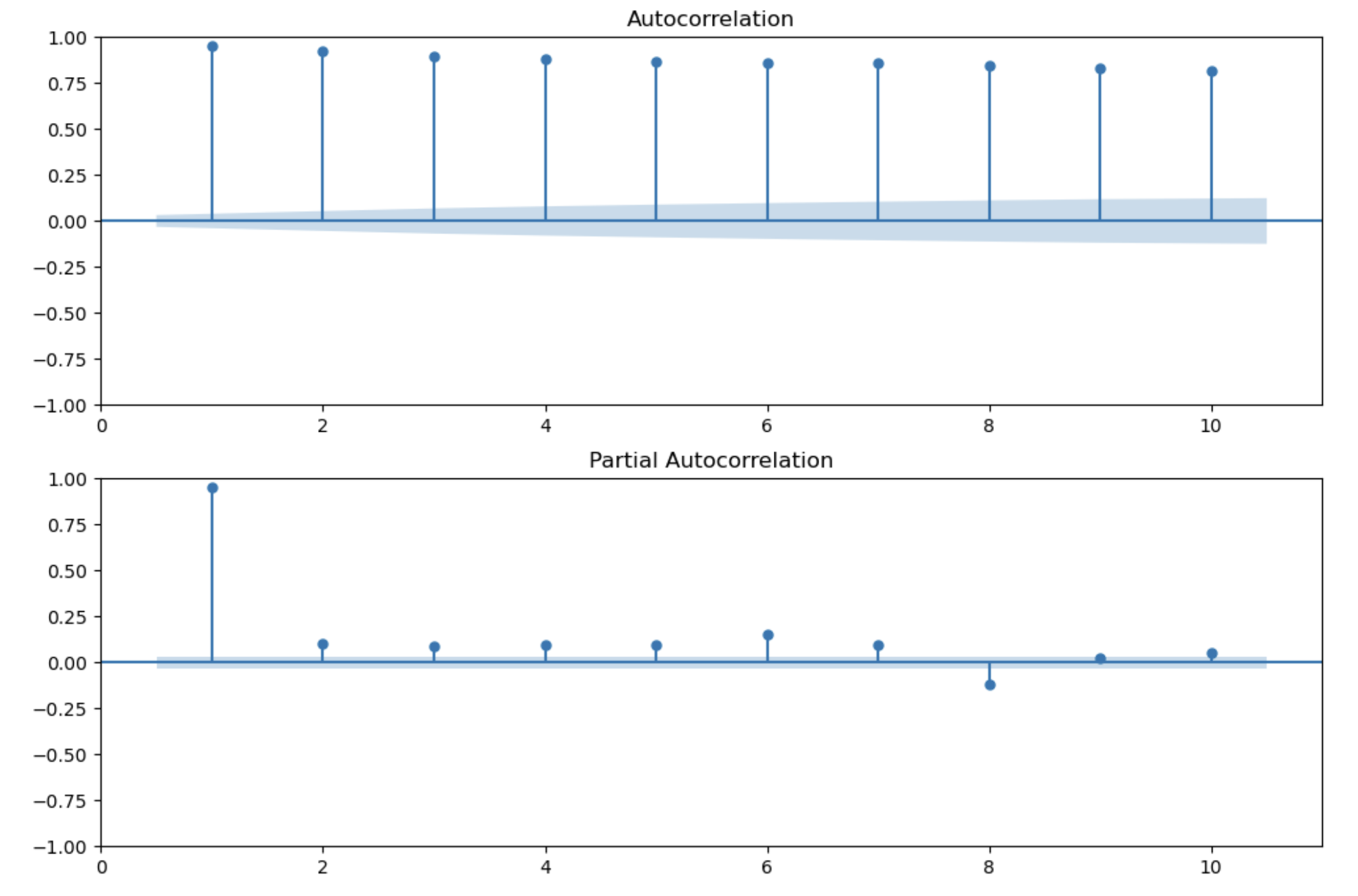}

The ADF test for first difference to try to change the time-series to stationary.
\hfill \break
\includegraphics[scale=0.4]{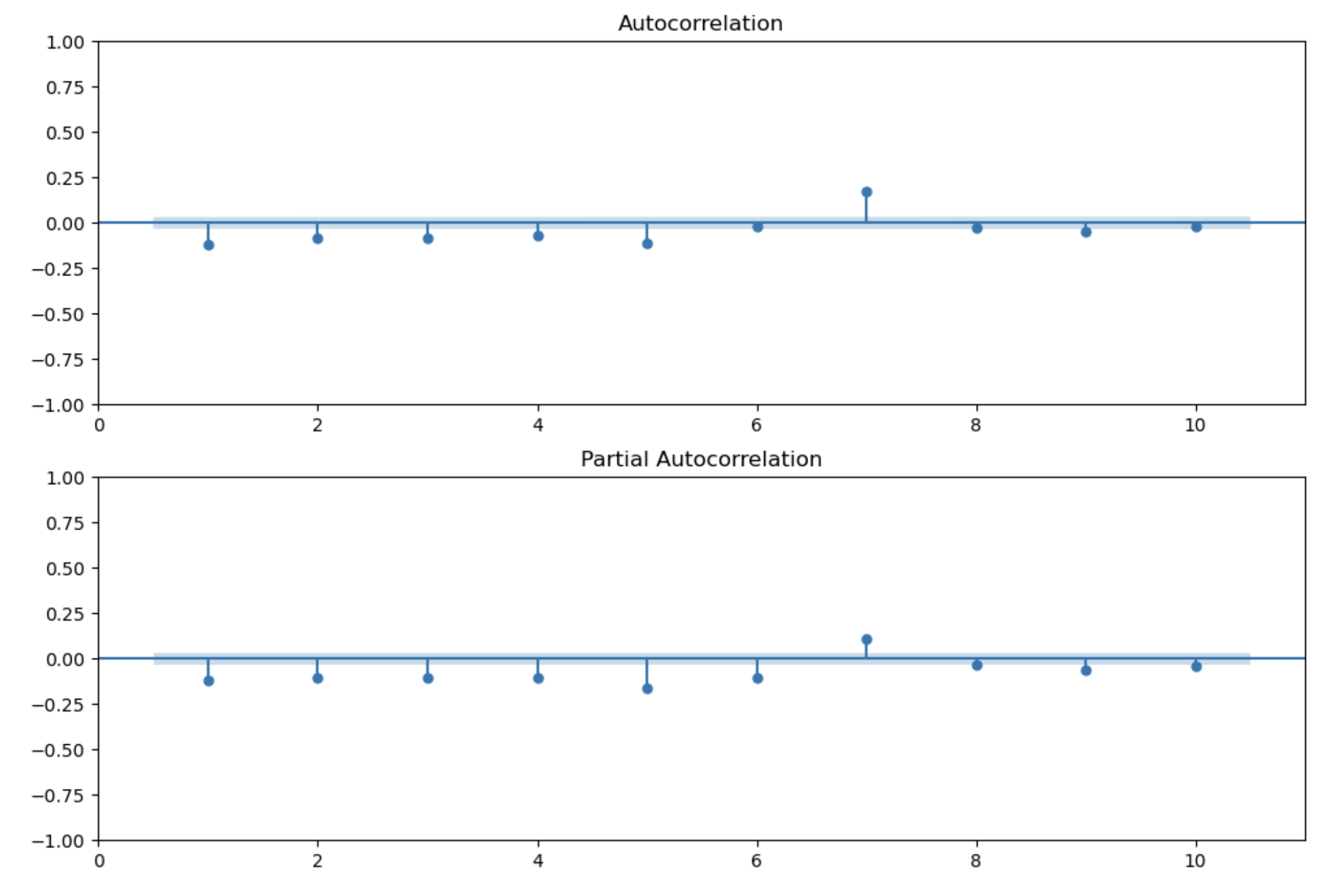}

\subsection{Median Imputation}
The Auto-Correlation Function and Partial Auto-Correlation Function Graph for the original dataset
\hfill \break
\includegraphics[scale=0.4]{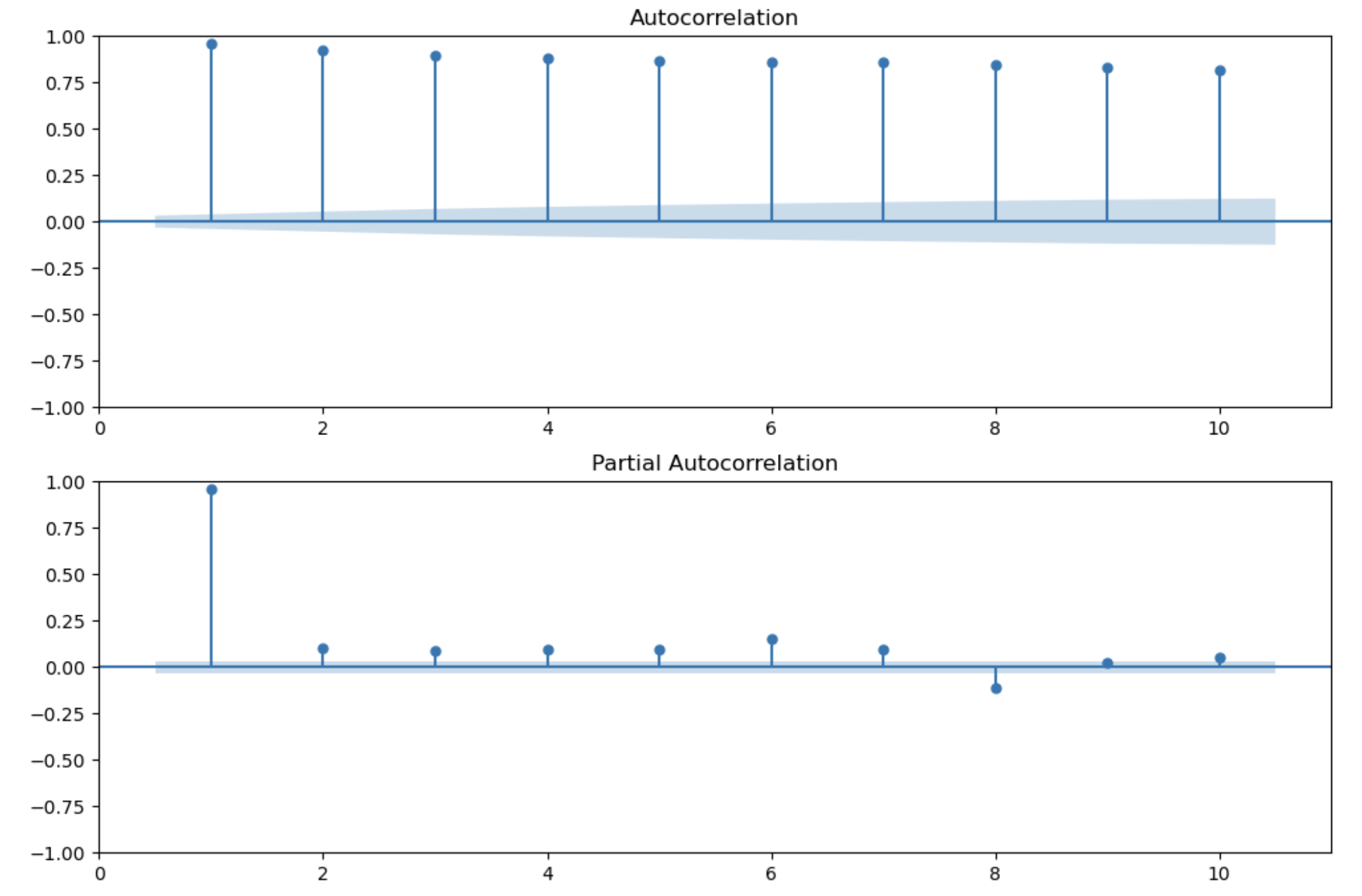}

The ADF test for first difference to try to change the time-series to stationary.
\hfill \break
\includegraphics[scale=0.4]{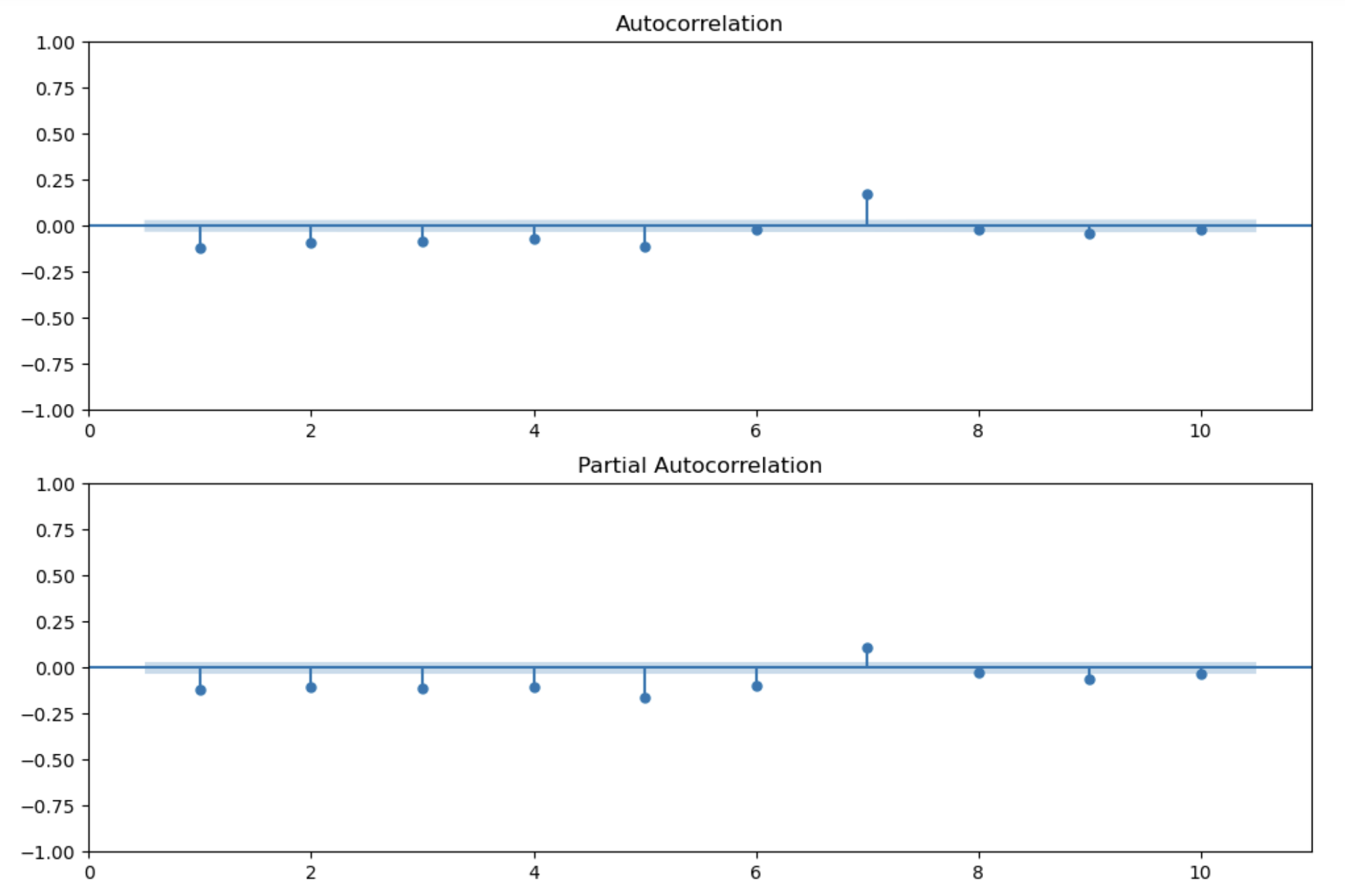}

\subsection{Mode Imputation}
The Auto-Correlation Function and Partial Auto-Correlation Function Graph for the original dataset
\hfill \break
\includegraphics[scale=0.4]{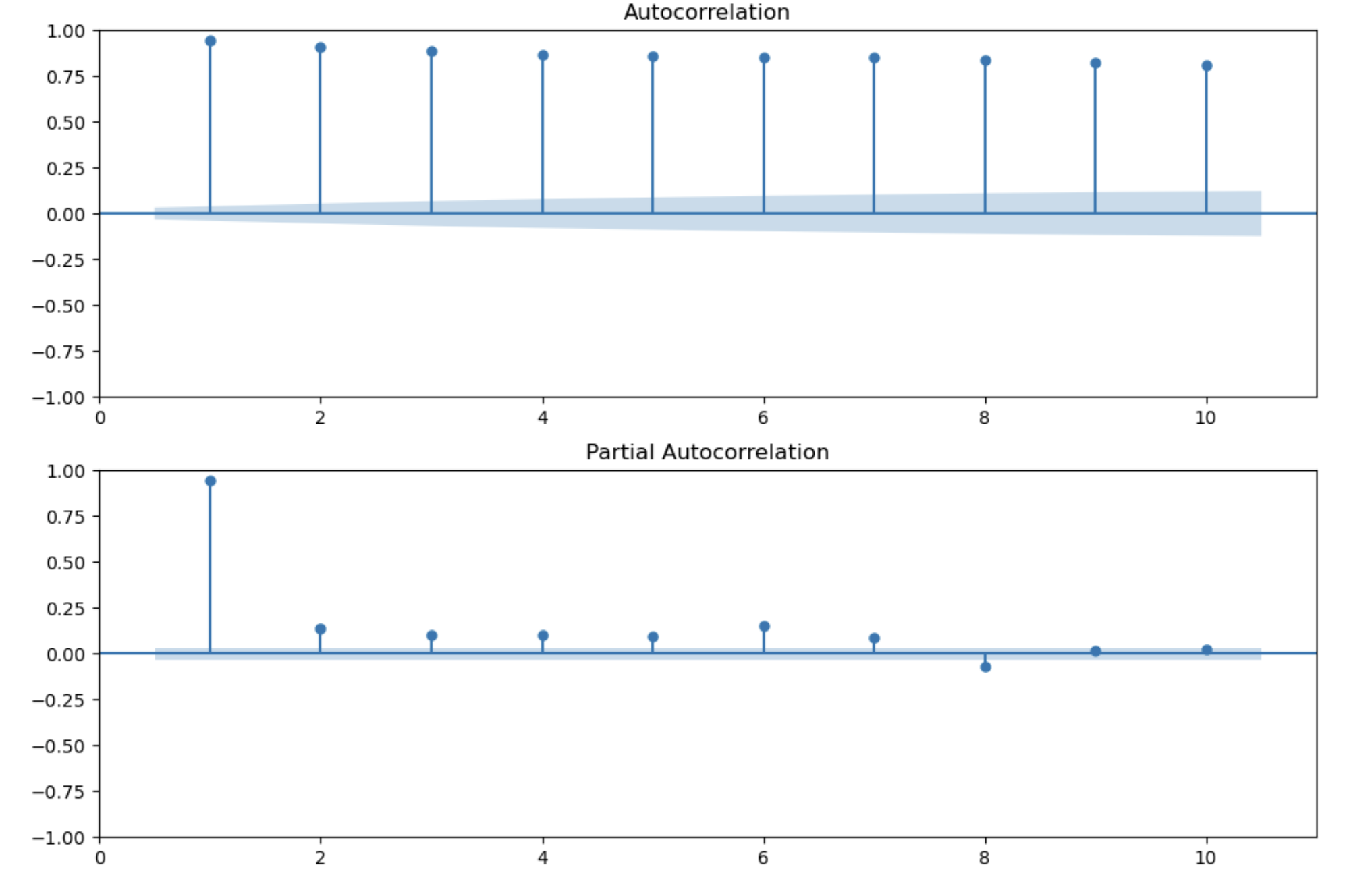}

The ADF test for first difference to try to change the time-series to stationary.
\hfill \break
\includegraphics[scale=0.4]{images/c5_9.png}

\subsection{Interpolation Imputation}
The Auto-Correlation Function and Partial Auto-Correlation Function Graph for the original dataset
\hfill \break
\includegraphics[scale=0.4]{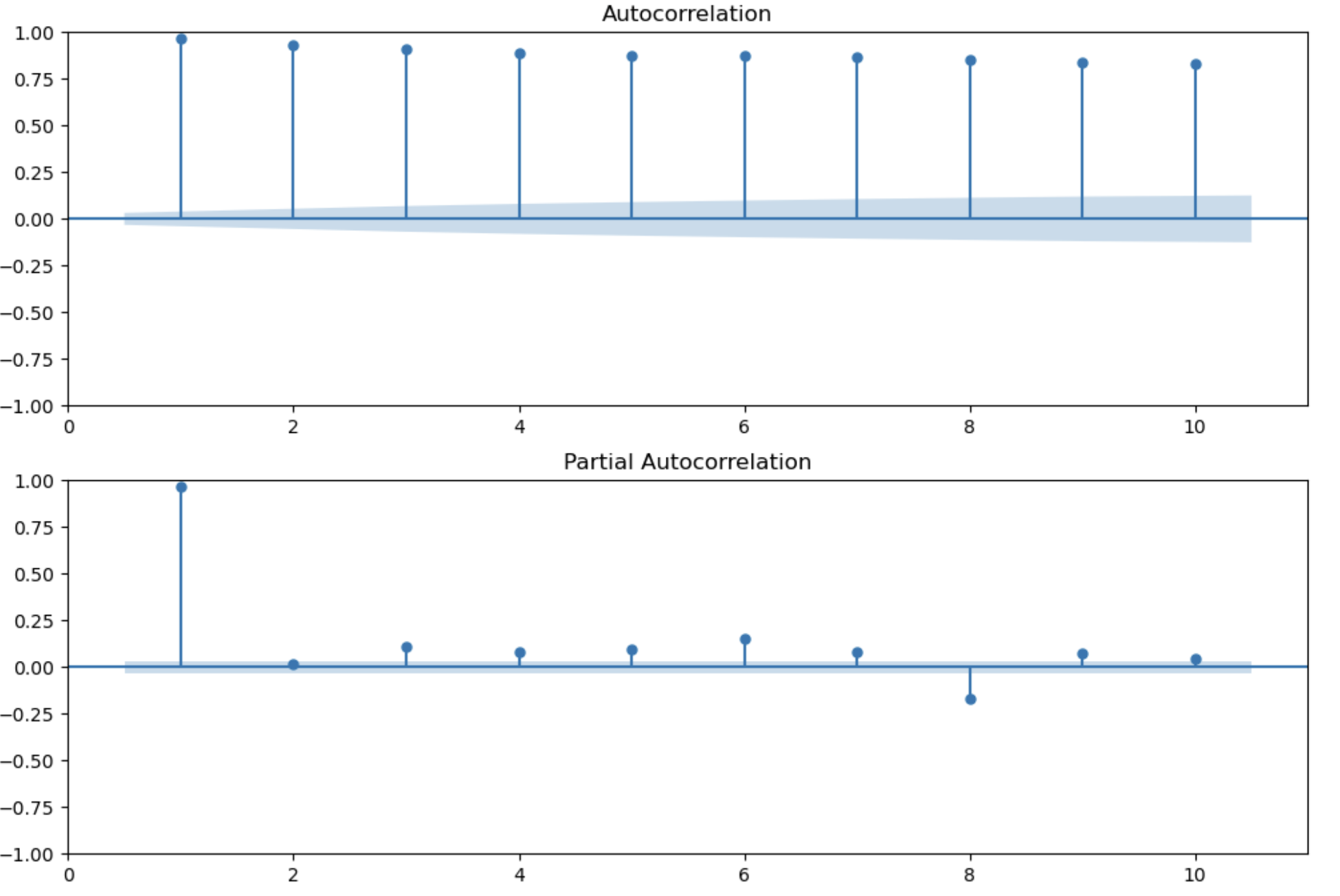}

The ADF test for first difference to try to change the time-series to stationary.
\hfill \break
\includegraphics[scale=0.4]{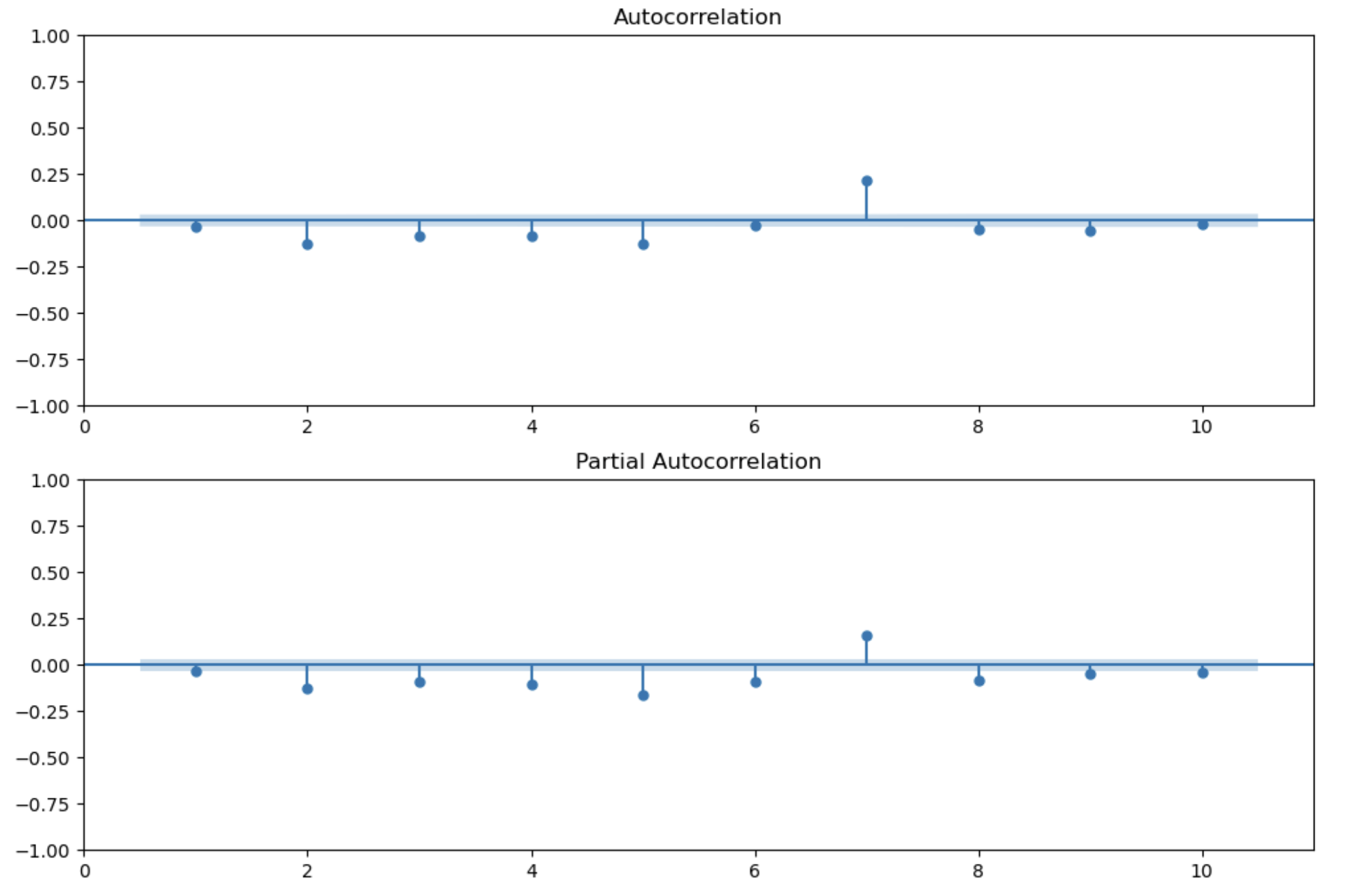}

\subsection{Auto Regression and Moving Average model}
In time series forecasting, the autoregressive moving average model of order $(p, q)$, denoted as ARMA($p, q$), is a popular approach. The ARMA($p, q$) model combines the autoregressive (AR) model of order $p$ and the moving average (MA) model of order $q$. The ARMA($p, q$) model assumes that the value of the time series at a given point is linearly dependent on the previous $p$ values of the series and the previous $q$ error terms. The formula for the ARMA($p, q$) model is as follows:

\[ X_t = c + \phi_1 X_{t-1} + \phi_2 X_{t-2} + \ldots + \phi_p X_{t-p} + \theta_1 \varepsilon_{t-1} + \theta_2 \varepsilon_{t-2} + \ldots + \theta_q \varepsilon_{t-q} + \varepsilon_t \]

In this formula:
\begin{itemize}
  \item $X_t$ represents the value of the time series at time $t$.
  \item $c$ is the intercept or constant term.
  \item $\phi_1, \phi_2, \ldots, \phi_p$ are the coefficients of the autoregressive terms that capture the relationship between the current and previous values.
  \item $X_{t-1}, X_{t-2}, \ldots, X_{t-p}$ represent the lagged values of the time series.
  \item $\theta_1, \theta_2, \ldots, \theta_q$ are the coefficients of the moving average terms that capture the relationship between the current value and the previous error terms.
  \item $\varepsilon_{t-1}, \varepsilon_{t-2}, \ldots, \varepsilon_{t-q}$ represent the lagged error terms of the time series.
  \item $\varepsilon_t$ is the error term at time $t$, which represents the random fluctuations or noise in the series.
\end{itemize}

To estimate the parameters ($\phi_1, \phi_2, \ldots, \phi_p, \theta_1, \theta_2, \ldots, \theta_q$) and the intercept ($c$) of the ARMA($p, q$) model, various estimation techniques can be used, such as maximum likelihood estimation.

Once the parameters are estimated, the ARMA($p, q$) model can be used for forecasting by substituting the lagged values and lagged error terms of the time series into the formula to predict future values.

Note that the ARMA($p, q$) model assumes stationarity of the time series, and it is a flexible model that can capture both autoregressive and moving average components in the data.

\subsection{Seasonal Auto-Regressive Models}
The Seasonal Autoregressive Integrated Moving Average (SARIMA) model is a time series forecasting model that extends the Autoregressive Integrated Moving Average (ARIMA) model to account for seasonality. SARIMA combines the components of ARIMA with seasonal differencing and seasonal autoregressive and moving average terms.

The SARIMA(p, d, q)(P, D, Q, s) model is defined by the following equations:

Autoregressive (AR) component:
AR(p): \(Y_t = \phi_1 Y_{t-1} + \phi_2 Y_{t-2} + \ldots + \phi_p Y_{t-p} + \varepsilon_t\)

Integrated (I) component:
I(d): \(Y'_t = (1-B)^d Y_t\), where \(B\) is the backshift operator (\(B Y_t = Y_{t-1}\))

Moving Average (MA) component:
MA(q): \(Y_t = \theta_1 \varepsilon_{t-1} + \theta_2 \varepsilon_{t-2} + \ldots + \theta_q \varepsilon_{t-q} + \varepsilon_t\)

Seasonal Autoregressive (SAR) component:
SAR(P): \(Y_t = \Phi_1 Y_{t-s} + \Phi_2 Y_{t-2s} + \ldots + \Phi_P Y_{t-Ps} + \varepsilon_t\)

Seasonal Moving Average (SMA) component:
SMA(Q): \(Y_t = \Theta_1 \varepsilon_{t-s} + \Theta_2 \varepsilon_{t-2s} + \ldots + \Theta_Q \varepsilon_{t-Qs} + \varepsilon_t\)

where:
\(Y_t\) is the observed time series at time \(t\)
\(\varepsilon_t\) is the error term (also known as the residual) at time \(t\)
\(p, d, q\) are the non-seasonal AR, I, MA orders, respectively
\(P, D, Q\) are the seasonal SAR, I, SMA orders, respectively
\(s\) is the seasonal period or frequency (e.g., 12 for monthly data, 4 for quarterly data, etc.)
\(\phi_1, \phi_2, \ldots, \phi_p\) are the non-seasonal autoregressive coefficients
\(\theta_1, \theta_2, \ldots, \theta_q\) are the non-seasonal moving average coefficients
\(\Phi_1, \Phi_2, \ldots, \Phi_P\) are the seasonal autoregressive coefficients
\(\Theta_1, \Theta_2, \ldots, \Theta_Q\) are the seasonal moving average coefficients

\begin{table*}[htbp]
\caption{ARIMA Model Comparison Results for dropna}
\centering
\resizebox{\dimexpr\textwidth+0.5px\relax}{!}{%
\begin{tabular}{|c|c|c|c|c|c|}
\hline
Models & Order & test\_MAPE & train\_MAPE & AIC & BIC \\
\hline
AR/MA models & 1,0,0 & 19.483 & 26.125 & 50534.716 & 50553.290 \\
             & 2,0,0 & 19.516 & 26.126 & 50534.362 & 50559.127 \\
             & 1,1,0 & 21.903 & 26.602 & 50575.065 & 50587.447 \\
             & 1,2,0 & 90.594 & 27.274 & 52837.437 & 52829.454 \\
             & 0,0,1 & 16.617 & 21.619 & 56254.968 & 56273.542 \\
             & 0,1,1 & 21.811 & 26.594 & 50572.554 & 50584.936 \\
             & 0,2,1 & 21.9037 & 26.602 & 50578.459 & 50590.841 \\
\hline
ARMA models  & 8,0,8 & 18.157 & 26.332 & 50067.375 & 50178.816 \\
             & 8,1,8 & 18.814 & 26.453 & 50148.885 & 50254.131 \\
             & 9,0,7 & 18.201 & 26.226 & 50067.729 & 50179.170 \\
             & 9,1,7 & 18.332 & 26.453 & 50055.038 & 50160.283 \\
             & 8,0,9 & 18.135 & 26.280 & 50068.705 & 50186.337 \\
             & 8,1,9 & 18.274 & 26.464 & 50059.330 & 50170.766 \\
\hline
auto-arima   & 5,1,3 & 18.844 & 26.454 & 50190.242 & 50245.961 \\
\hline

\end{tabular}
}
\end{table*}

\begin{table*}[htbp]
\caption{ARIMA model Comparison Results for Mean Imputation}
\centering
\resizebox{\dimexpr\textwidth+10px\relax}{!}{%
\begin{tabular}{|c|c|c|c|c|c|}
\hline
Models & Order & test\_MAPE & train\_MAPE & AIC & BIC \\
\hline
AR/MA models & 1,0,0 & 19.186 & 25.749 & 52184.042 & 52202.676 \\
             & 2,0,0 & 19.352 & 25.756 & 52144.192 & 52169.037 \\
             & 1,1,0 & 21.628 & 26.276 & 52188.319 & 52200.741 \\
             & 1,2,0 & 86.037 & 27.046 & 54529.566 & 54542.004 \\
             & 0,0,1 & 16.616 & 21.290 & 57600.735 & 57619.368 \\
             & 0,1,1 & 21.322 & 26.248 & 52171.044 & 52183.466 \\
             & 0,2,1 & 21.628 & 26.276 & 52243.646 & 52256.067 \\
\hline
ARMA models  & 9,0,8 & 18.266 & 26.027 & 51635.298 & 51753.311 \\
             & 9,1,8 & 18.069 & 26.135 & 51687.745 & 51799.541 \\
             & 8,0,9 & 18.400 & 25.965 & 51663.391 & 51781.404 \\
             & 8,1,9 & 18.366 & 26.173 & 51695.921 & 51807.718 \\
             & 8,0,8 & 18.518 & 25.977 & 51670.186 & 51781.988 \\
             & 8,1,8 & 18.324 & 26.093 & 51637.395 & 51742.981 \\
\hline
auto-arima   & 5,1,3 & 18.563 & 26.085 & 51820.469 & 51876.368 \\
\hline
\end{tabular}%
}
\end{table*}

\begin{table*}[htbp]
\caption{ARIMA model Comparison Results for Median Imputation}
\centering
\resizebox{\textwidth}{!}{%
\begin{tabular}{|c|c|c|c|c|c|}
\hline
Models & Order & test\_MAPE & train\_MAPE & AIC & BIC \\
\hline
AR/MA models & 1,0,0 & 19.203 & 25.740 & 52180.550 & 52199.184 \\
             & 2,0,0 & 19.363 & 25.748 & 52142.980 & 52167.825 \\
             & 1,1,0 & 21.636 & 26.268 & 52187.390 & 52199.812 \\
             & 1,2,0 & 86.201 & 27.038 & 54527.913 & 54540.351 \\
             & 0,0,1 & 16.646 & 21.271 & 57600.941 & 57619.575 \\
             & 0,1,1 & 21.336 & 26.241 & 52170.583 & 52183.005 \\
             & 0,2,1 & 21.636 & 26.268 & 52239.944 & 52252.365 \\
\hline
ARMA models  & 9,0,8 & 18.288 & 26.032 & 51635.740 & 51753.753 \\
             & 9,1,8 & 18.115 & 26.121 & 51681.885 & 51793.682 \\
             & 8,0,8 & 18.506 & 25.958 & 51674.400 & 51786.202 \\
             & 8,1,8 & 18.421 & 26.104 & 51641.077 & 51746.663 \\
             & 8,0,9 & 18.407 & 25.939 & 51668.680 & 51786.693 \\
             & 8,1,9 & 18.502 & 26.152 & 51701.538 & 51813.335 \\
\hline
auto-arima   & 5,1,3 & 18.676 & 26.098 & 51805.777 & 51861.675 \\
\hline
\end{tabular}%
}
\end{table*}

\begin{table*}[htbp]
\caption{ARIMA model Comparison Results for Median Imputation}
\centering
\resizebox{\textwidth}{!}{%
\begin{tabular}{|c|c|c|c|c|c|}
\hline
Models & Order & test\_MAPE & train\_MAPE & AIC & BIC \\
\hline
AR/MA models & 1,0,0 & 19.025 & 25.966 & 52851.514 & 52870.148 \\
             & 2,0,0 & 19.256 & 25.977 & 52780.739 & 52805.583 \\
             & 1,1,0 & 21.531 & 26.566 & 52831.433 & 52843.854 \\
             & 1,2,0 & 85.088 & 27.498 & 55261.639 & 55274.078 \\
             & 0,0,1 & 16.796 & 21.587 & 57849.604 & 57868.238 \\
             & 0,1,1 & 21.013 & 26.513 & 52795.815 & 52808.237 \\
             & 0,2,1 & 21.531 & 26.566 & 52925.747 & 52938.168 \\
\hline
ARMA models  & 9,0,8 & 18.004 & 26.273 & 52357.307 & 52475.320 \\
             & 9,1,8 & 22.428 & 26.522 & 52691.287 & 52803.084 \\
             & 9,0,9 & 18.237 & 26.157 & 52376.970 & 52501.194 \\
             & 9,1,9 & 16.651 & 26.512 & 52356.583 & 52474.591 \\
             & 8,0,9 & 18.343 & 26.220 & 52412.001 & 52530.014 \\
             & 8,1,9 & 18.698 & 26.342 & 52578.576 & 52690.373 \\
\hline
auto-arima & 3,1,4 & 18.401 & 26.340 & 52506.970 & 52562.869 \\
\hline
\end{tabular}%
}
\end{table*}

\begin{table*}[htbp]
\caption{ARIMA model Comparison Results for Linear Interpolation Imputation}
\centering
\resizebox{\textwidth}{!}{%
\begin{tabular}{|c|c|c|c|c|c|}
\hline
Models & Order & test\_MAPE & train\_MAPE & AIC & BIC \\
\hline
AR/MA models & 1,0,0 & 19.552 & 26.051 & 51429.997 & 51448.631 \\
             & 2,0,0 & 19.573 & 26.051 & 51430.941 & 51455.786 \\
             & 1,1,0 & 21.936 & 26.516 & 51472.149 & 51484.571 \\
             & 1,2,0 & 90.903 & 27.171 & 53740.953 & 53753.391 \\
             & 0,0,1 & 16.678 & 21.549 & 57337.414 & 57356.048 \\
             & 0,1,1 & 21.870 & 26.510 & 51470.598 & 51483.020 \\
             & 0,2,1 & 21.936 & 26.516 & 51473.203 & 51485.624 \\
\hline
ARMA models  & 8,0,8 & 18.409 & 26.271 & 50801.251 & 50913.053 \\
             & 8,1,8 & 18.683 & 26.385 & 50838.995 & 50944.581 \\
             & 9,0,8 & 18.223 & 26.175 & 50814.366 & 50932.379 \\
             & 9,1,8 & 18.440 & 26.385 & 50807.144 & 50918.941 \\
             & 9,0,7 & 18.157 & 26.114 & 50836.848 & 50948.650 \\
             & 9,1,7 & 19.432 & 26.363 & 50840.746 & 50946.332 \\
\hline
auto-arima   & 5,1,4 & 19.114 & 26.389 & 51004.208 & 51066.318 \\
\hline
\end{tabular}%
}
\end{table*}

\begin{table*}[htbp]
\caption{SARIMA model Comparison Results for dropna}
\centering
\resizebox{\textwidth}{!}{%
\begin{tabular}{|c|c|c|c|c|c|}
\hline
order & Seasonal-order & train\_MAPE & test\_MAPE & AIC & BIC \\
\hline
(1, 0, 0) & (3,0,6,7) & 26.673 & 20.554 & 50233.507 & 50301.610 \\
(0,0,0) & (1,0,1,7) & 26.301 & 17.40 & 54957.193 & 54975.766 \\
(0,0,0) & (1,1,1,7) & 26.461 & 16.882 & 54823.939 & 54842.507 \\
(0,0,0) & (3,0,6,7) & 25.505 & 15.670 & 54736.140 & 54798.052 \\
(0,0,0) & (3,1,6,7) & 26.826 & 15.854 & 54735.359 & 54797.251 \\
\hline
\end{tabular}%
}
\end{table*}

\begin{table*}[htbp]
\caption{ARIMA model Comparison Results for mean Imputation}
\centering
\resizebox{\textwidth}{!}{%
\begin{tabular}{|c|c|c|c|c|c|}
\hline
order & Seasonal-order & train\_MAPE & test\_MAPE & AIC & BIC \\
\hline
(1, 0, 0) & (6,0,2,7) & 25.884 & 17.775 & 51748.896 & 51811.008 \\
(0,0,0) & (1,0,1,7) & 25.973 & 17.382 & 56160.584 & 56179.218 \\
(0,0,0) & (1,1,1,7) & 26.141 & 16.831 & 56026.752 & 56045.380 \\
(0,0,0) & (6,0,2,7) & 25.355 & 15.398 & 55932.485 & 55988.386 \\
(0,0,0) & (6,1,2,7) & 26.404 & 16.173 & 55954.966 & 56010.850 \\
\hline
\end{tabular}%
}
\end{table*}

\begin{table*}[htbp]
\caption{SARIMA model Comparison Results for Median Imputation}
\centering
\resizebox{\textwidth}{!}{%
\begin{tabular}{|c|c|c|c|c|c|}
\hline
order & Seasonal-order & test\_MAPE & train\_MAPE & AIC & BIC \\
\hline
(1, 0, 0) & (3,0,6,7) & 20.491 & 26.316 & 51825.927 & 51894.250 \\
(0,0,0) & (1,0,1,7) & 17.379 & 25.965 & 56152.138 & 56170.772 \\
(0,0,0) & (1,1,1,7) & 16.830 & 26.132 & 56018.390 & 56037.018 \\
(0,0,0) & (3,0,6,7) & 15.717 & 25.119 & 55924.073 & 55986.185 \\
(0,0,0) & (3,1,6,7) & 15.950 & 26.439 &  55951.882 & 56013.975 \\
\hline

\end{tabular}%
}
\end{table*}

\begin{table*}[htbp]
\caption{SARIMA model Comparison Results for Mode Imputation}
\centering
\resizebox{\textwidth}{!}{%
\begin{tabular}{|c|c|c|c|c|c|}
\hline
order & Seasonal-order & test\_MAPE & train\_MAPE & AIC & BIC \\
\hline
(1, 0, 0) & (3,0,6,7) & 20.554 & 26.673 & 50233.507 & 50301.610 \\
(0,0,0) & (1,0,1,7) & 17.407 & 26.301 & 54957.193 & 54975.766 \\
(0,0,0) & (1,1,1,7) & 16.882 & 26.461 & 54823.939 & 54842.507 \\
(0,0,0) & (3,0,6,7) & 15.670 & 25.505 & 54736.140 & 54798.052 \\
(0,0,0) & (3,1,6,7) & 15.854 & 26.826 & 54735.359 & 54797.251 \\
\hline

\end{tabular}%
}
\end{table*}

\begin{table*}[htbp]
\caption{SARIMA model Comparison Results for Linear Interpolation Impuation}
\centering
\resizebox{\textwidth}{!}{%
\begin{tabular}{|c|c|c|c|c|c|}
\hline
order & Seasonal-order & test\_MAPE & train\_MAPE & AIC & BIC \\
\hline
(1, 0, 0) & (6,0,3,7) & 19.905 & 26.556 & 50917.697 & 50986.020 \\
(0,0,0) & (1,0,1,7) & 17.401 & 26.212 & 55984.215 & 56002.849 \\
(0,0,0) & (1,1,1,7) & 16.871 & 26.372 & 55851.605 & 55870.233 \\
(0,0,0) & (6,0,3,7) & 15.304 & 26.425 & 55773.403 & 55835.515 \\
(0,0,0) & (6,1,3,7) & 15.210 & 26.442 & 55627.806 & 55689.899 \\
\hline

\end{tabular}%
}
\end{table*}

\pagebreak

\section{Results and Conclusion}

Table 1 to 11 depicts the results of the Classical Time-series Forecasting methods. SARIMA(0,0,0)(6,1,3,7) is the best model which is provided by the MAPE scores of the test data. As the Base models concluded we try to integrate more data into the PM Gati-Shakti Scheme to validate the case study given in the section of Introduction. The future work also focuses on using Reinforcement Learning for model selection for much larger data with integrated ministries in the Union Territory of Delhi.

In conclusion, our study highlights the importance of integrating more data into the PM Gati-Shakti Scheme in order to validate the findings presented in the Introduction section. The base models provide a preliminary understanding of the scheme's potential, but further data incorporation is crucial for robust conclusions. By expanding the scope of our analysis to encompass a wider range of variables and factors, we can enhance the accuracy and reliability of the case study.

Furthermore, our future work will focus on employing Reinforcement Learning techniques for model selection. This approach is particularly relevant when dealing with a larger dataset that integrates ministries within the Union Territory of Delhi. Reinforcement Learning algorithms can effectively evaluate and select the most suitable models by considering the complex interactions and dependencies between different variables. By leveraging the power of machine learning and advanced analytics, we can make informed decisions that lead to better outcomes and enhanced efficiency within the PM Gati-Shakti Scheme.

In summary, our research emphasizes the need for data integration and the application of Reinforcement Learning in the context of the PM Gati-Shakti Scheme. These steps will contribute to a more comprehensive understanding of the scheme's impact and enable evidence-based decision-making for the integration of ministries in the Union Territory of Delhi. By continuously improving our analytical approaches, we can enhance the effectiveness of the scheme and drive positive socio-economic outcomes.

\section*{Acknowledgments}
This project was supported by National Institute of Technology, Calicut a center of PM-Gati-Shakti Scheme initiated by National Institute of Industrial Engineering.


\end{document}